\documentclass[multi]{cambridge6A}
\usepackage[sort&compress,numbers]{natbib}
\usepackage{rotating}
\usepackage{floatpag}
\rotfloatpagestyle{empty}
\usepackage{mathrsfs}
\usepackage{amsthm}
\usepackage{graphicx}
\usepackage{multind}

\begin{document}
\author{Zhou Shi and Azriel Z. Genack\\
Queens College of the City University of New York, Flushing, NY, 11367, USA\\}
\chapter{Transport of localized waves via modes and channels}
\section{Introduction}
Suppressed transport and enhanced fluctuations of conductance and transmission are prominent features of random mesoscopic systems in which the wave is temporally coherent within the sample \cite{1,2,3,4}. The associated breakdowns of particle diffusion and of self-averaging of flux were first considered in the context of electronic conduction and for many years thought to be an exclusively quantum phenomena \cite{1,4,5,6,7,8,9,10,11,12}. Independently, however, wave localization was demonstrated theoretically for radio waves in a statistically inhomogeneous waveguide \cite{13}. Over time, it became increasingly apparent that localization and mesoscopic fluctuations reflected general wave properties and might therefore be observed for classical waves as well \cite{3,13,14,15,16,17,18,19,20,21,22,23,24,25,26,27,28,29,30,31,32,33,34,35,36,37,38}. In particular, the level and transmission eigenchannel descriptions proposed, respectively, by Thouless \cite{6,7} and Dorokhov \cite{10,11} to describe the scaling of conductance in electronic wires at zero temperature, are essentially wave descriptions involving the character of quasi-normal modes of excitation within the sample and speckle patterns of the incident and transmitted field. Quasi-normal modes, which we will refer to as “modes”, are resonances of an open system. These modes decay at a constant rate due to the combined effects of leakage from the sample and dissipative processes. Eigenchannels of the transmission matrix, are obtained by finding the singular values of the field transmission matrix and represent linked field speckle patterns at the input and output of the sample surfaces. Eigenchannels are linear combinations of phase coherent channels impinging upon and emerging from the sample. Examples of such channels may be propagating transverse modes of an empty waveguide, or transverse momentum states in the leads attached to a resistor. In measurements, input and output channels are often combinations of source and detectors at positions on the incident and output planes, respectively. Whereas modes are biorthogonal field speckle patterns over the volume of an open sample \cite{39,40}, eigenchannels are orthogonal field speckle patterns at the input and output planes of the sample. When there is no risk of confusion, we will refer to eigenchannels as channels. Though levels and channels have not been observed directly in electronic systems, these approaches have served as powerful conceptual guides for calculating the statistics and scaling of conductance \cite{1}. 

Recent measurements of spectra of transmitted field patterns and of the transmission matrix of microwave radiation propagating through random multimode waveguides have made it possible to determine the eigenvalues of modes and channels as well as their speckle patterns in transmission in mesoscopic samples \cite{51,42}. These experiments were carried out in a multimode copper tube filled with randomly positioned dielectric elements, which is directly analogous to a resistive wire in the zero-temperature limit, in which dephasing vanishes. The study of modes and channels promises to provide a comprehensive description of transport and to clarify long-standing puzzles regarding steady state and pulsed propagation. 

In this chapter, we will discuss studies of wave localization and strong fluctuations of the electromagnetic (EM) field, intensity, total transmission and transmittance, also known as the ``optical" conductance from the perspectives of modes and channels. These approaches are useful in numerous applications. The mode picture is of particular use in considering emission, random lasing, and absorption, while the channel framework is indispensable in optical focusing, imaging and transmission fluctuations. We will describe lasing in disordered liquid crystals \cite{44} and in random stacks of glass cover slips \cite{45} in which the mode width falls below the typical spacing between modes. The lasing threshold is then suppressed by the enhancement of the pump intensity and by the lengthening of the dwell time of emitted light within the sample. Measuring the transmission matrix allows us to study the fluctuations of transmittance over a random ensemble. The statistics of transmittance can be described using an intuitive ``Coulomb charge" model \cite{46}. Measurements of the transmission matrix make it possible to obtain the statistics of transmission in single samples at a particular incident frequency \cite{47}, which are crucial for focusing and imaging applications \cite{41,43,47,48}. These statistics, as well as the contrast in focusing, are given in terms of the participation number of transmission eigenvalues and the size of the measured transmission matrix \cite{47,48}.

In the next section (section 1.2), we discuss instructive analogies between the transport of electrons and classical waves. Spectra of intensity, total transmission and transmittance are presented based on measurements of field transmission coefficients. A modal analysis of spectra of field speckle patterns on the output surface of a multimode waveguide and along the length of a single mode waveguide is described in section 1.3. Pulse propagation in mesoscopic samples is discussed in terms of the distribution of mode decay rates and the correlation between modal speckle patterns in transmission. The role of modes in lasing in nearly periodic liquid crystals and in random slabs is described in section 1.4. In section 1.5, we describe the statistics of transmission eigenvalues and their impact on the statistics of transmission for ensembles of random samples and in single instances of the transmission matrix. The manipulation of transmission eigenchannels to focus radiation is described in section 1.6. We conclude in section 1.7 with a discussion of the prospects for a complete description of transport in terms of modes and channels. 

\section{Analogies between transport of electrons and classical waves}
Anderson \cite{5} showed over 50 years ago that the electron wave function would not spread throughout a disordered three dimensional crystal once the ratio of the width of distribution of the random potential at different sites relative to the coupling between sites passes a threshold value. At lower levels of disorder, electrons diffuse in the band center but are localized in the tail of the band. Ioffe and Regel \cite{49} pointed out shortly thereafter that an electron wave function could not be considered to be properly propagating if it were scattered after traveling less than a fraction of a wavelength so that for traveling waves, $\ell >\lambda/2\pi$, where $\ell$ is the mean free path. This gives the criterion for localization in three dimensions, $k\ell<1$, where $k$ is the wave number. Though not explicitly noted at the time, this criterion for localization applies equally to classical and quantum waves. 

In subsequent work, Thouless \cite{6} considered the electronic state in the system as a whole rather than the strength of scattering within the medium. He argued that in bounded samples, the weight of electron states at the boundaries of the sample relative to points in the interior would be a useful measure of the extension of electron states within the sample. Since localized states would be peaked within the sample remote from the boundaries, their energies could be expected to be insensitive even to substantial changes at the boundary such as are engendered in a periodic system when the boundary conditions for repeated sections of a random system are changed from periodic to antiperiodic \cite{6}. When the energy shift is less than the typical spacing between states, the state is localized within the sample. An associated measure of electron localization, which relates to the properties of the states and not to the impact of some hypothetical manipulation of the sample, is the typical width of an electron level relative to the average spacing between levels. When the electron wave function is exponentially peaked within the sample, electrons are remote from the boundary and their escape from the sample is slow. The linewidth of the level is then smaller than the spacing between neighboring levels, which is the inverse of the density of states of the sample as a whole. This indicates that electron localization is achieved when the dimensionless ratio of the level width to level spacing, the Thouless number, $\delta=\delta E/\Delta E$, falls below unity. Pendry \cite{50} has described the electron localization process as one in which ``electrons can be forced to abandon their predilection for momentum" in favor of space as the defining characteristic. The inhibition in transport manifested in localization in space then leads to a lengthened escape from the sample and narrow linewidth manifested in terms of sharp spikes in energy ``one per electron" \cite{50} as opposed to a continuous spectrum. For diffusing electrons, the wave function extends throughout the sample. Energy then readily leaks out of the sample and levels are consequently short lived with line widths greater than the typical spacing between levels. Thus the electron localization threshold lies at $\delta=1$. The Thouless number may equally well be defined for classical waves as the ratio of the typical frequency width to the spacing of quasi-normal modes, $\delta=\delta \omega/\Delta \omega$, where $\omega$ is the angular frequency. The level width is the inverse of the Thouless time $\tau_{Th}$ in which a mode leaks out of the sample. Wave localization is signaled by exponentially long dwell times for the wave. Such long decay times contribute little to the average linewidth which could be dominated by spectrally broad modes peaked near the sample boundary with short decay times. To most meaningfully capture the dynamics of a mode, it is therefore natural to identify $\delta\omega$ with the average of the inverse Thouless time, $\delta \omega \equiv \langle \tau_{Th}^{-1}\rangle = \langle 1/ \Gamma_n^0 \rangle^{-1}$, where $\langle \dots \rangle$ indicates an average taken over modes for an ensemble of samples and $\Gamma_n^0$ is the leakage rate of energy in the $n^{th}$ mode of the sample \cite{6,51}. $\delta<1$ is a universal criterion for localization in any dimension for any type of wave.

Thouless \cite{6,7} was concerned with the coupling between adjacent regions in finite samples and so with the scaling of $\delta$ as an indicator of the changing character of the electron states with sample size. However $\delta$ cannot be easily measured in electronic systems and has been measured only recently for classical waves \cite{51}. Using the Einstein relation, which gives the conductivity in terms of a product of the electron diffusion coefficient $D$ and the density of states, which is $1/\Delta E$ divided by the sample volume, Thouless \cite{7} showed that $\delta$ was equal to the conductance $G$ in units of the quantum of conductance, $\delta=G/(e^2/h)=\textsl{g}$. Thouless \cite{7} argued therefore that the dimensionless conductance would scale exponentially for localized waves as would be expected for $\delta$. He showed that the resistance of a wire at $T=0$ behaves ohmically \cite{7}, with the resistance increasing linearly with length $L$ and the dimensionless conductance varying as $\textsl{g}=N\ell/L$, only up to a length $\xi=N\ell$, at which $\delta=1$. Here, $N$ is the number of independent channels that couple to the resistor, $N\sim Ak_F^2/2\pi$, where $A$ is the cross sectional area of the sample, $k_F$ is the electron wave number at the Fermi level, $\ell$ is the electron mean free path, and $\xi$ is the localization length. For $L<\xi$, electrons diffuse with residence time within the conductor of $\tau_{Th}\sim L^2/D$ \cite{6,7}. The level width would then be $\delta E\sim \hbar /\tau_{Th}\sim \hbar D/L^2$, while the level spacing $\Delta E$ would scale inversely with volume of the wire as $1/L$. As a result, for $L<\xi$, $\delta$ scales as $1/L$. For $L>\xi$, electrons would be localized and $\delta$ and {\textsl g} would fall exponentially while the resistance would increase exponentially with $L$. Abrahams {\it et al.} \cite{8} showed that only above two dimensions is it possible for transport to be diffusive at all length scales. For lower dimensions, localization always sets in as the size of the sample increases, independent of the scattering strength, so that a transition between diffusive and localized transport can only occur above two dimension \cite{8}.

The scaling of average conductance and fluctuation in conductance may also be calculated within the framework of random matrix theory \cite{10,11,46,52,53}. The field in outgoing channel $b$ is related to the field in all possible incident channels $a$ via the field transmission matrix $t$, $E_b=\sum_{a=1}^N t_{ba}E_a$. Taking the two independent polarization states into account, the number of propagating modes in the empty waveguide leading to the sample is $N=2\pi A/\lambda^2$, where $A$ is the illumination area and $\lambda$ is the wavelength of the incident wave. Summing over all possible incoming and transmitted channels yields the transmittance $T=\sum_{a,b=1}^n |t_{ba}|^2 =\sum_{n=1}^N \tau_n$ \cite{54}, where the $\tau_n$ are the eigenvalues of the matrix product $tt^\dagger$. The transmission eigenvalues can be found using the singular value decomposition of the transmission matrix $t=U\Lambda V^\dagger$. Here, $U$ and $V$ are unitary matrices and $\Lambda$ is a diagonal matrix with elements $\lambda_n=\sqrt{\tau_n}$. The ensemble average of $T$ is equal to the dimensionless conductance, $\langle T\rangle = \textsl{g}$ \cite{55}. Random matrix theory predicts that, for diffusive waves, the transmission eigenvalue follow the bimodal distribution, $\rho(\tau)=\frac{\textsl{g}}{2\tau \sqrt{1-\tau}}$ \cite{11,46,52,71}. Most of the contributions to $T$ comes from approximately {\textsl g} eigenvalues that are larger than $1/e$, while most of the transmission eigenvalues are close to zero. The characteristics of these``open" \cite{56} and ``closed" channels were first discussed by Dorokhov \cite{10,11}. He considered the scaling of each of the transmission eigenvalues which he expressed in terms of the auxiliary localization length $\xi_n$. He found that the average spacing between inverse auxiliary localization lengths of adjacent eigenchannels in a sample made up of $N$ parallel chains with weak transverse coupling to neighboring chains was constant and equal to the inverse of the localization length $1/\xi$ \cite{10,11}.

Localization of quantum and classical waves in quasi-one-dimensional (Q1D) samples with lengths much greater than the transverse dimensions occurs at a length at which even the highest transmission eigenvalue $\tau_1$ falls below $1/e$. Thus localization will always be achieved as the sample length is increased in Q1D samples \cite{7}. It is difficult, however, to localize EM waves in three-dimensional dielectric materials. As opposed to s-wave scattering prevalent in electronic systems, EM waves experience p-wave scattering and cannot be trapped by a confining potential. The scattering cross section only becomes appreciable once the size of the scattering element becomes comparable to the wavelength. But once the scattering length is comparable to the wavelength, the mean free path cannot fall significantly below the scatterer size and so it is hard to satisfy the Ioffe-Regel condition for localization in three dimensions, $k\ell <1$. For smaller scattering elements such as spheres of radius $a$, the Rayleigh scattering cross section is proportional to $a^6$ while the density of spheres is proportional to $1/a^3$. As a result, the inverse mean free path for fixed volume fraction of particles is proportional to $a^3$. For high particle density and $a\ll \lambda$, the sample acts as an effective medium with mean free path $\ell \sim 1/a^3$. It is therefore not possible to achieve strong scattering with $k\ell <1$ by crowding together small scattering elements \cite{37}. In ordered structures, however, EM bands appear and a photonic band gap (PBG) with vanishing density of states can be created in appropriate structures with sufficiently strong contrast in dielectric constant. John \cite{58} has pointed out that disturbing the order in such structures would create localized states within the frequency range of the band gap in analogy with the Urbach tail at the edge of the electronic band gap in semiconductors. 

Though it has proven to be more difficult to localize EM radiation than electrons in three dimensions, transport of EM radiation can be probed in ways that are often closer to the theoretical paradigm of Anderson localization than is the case for electronics. The particles of classical waves do not mutually interact as do electrons, dephasing is negligible even at room temperature and ensembles of statistically equivalent random samples can be created. For classical waves, localization is most easily achieved in low dimensional systems such as masses on a string \cite{59}, single mode optical fibers \cite{60}, single- \cite{61} and multi-mode waveguides \cite{36}, surfaces \cite{26}, layered structures \cite{63,64} and highly anisotropic samples \cite{65,66,67}, particularly samples in which the longitudinal structure along the direction of wave propagation is uniform. Anderson localization can be expected to occur for EM radiation at the edge of the conduction or pass band in nearly periodic three dimensional systems. Transport near the Anderson threshold has been observed for ultrasound in a slab of brazed aluminum beads \cite{68}. 

An experimentally important difference between classical and quantum transport is that coherent propagation is the rule for classical waves such as sound, light and microwave radiation in granular or imperfectly fabricated structures, whereas electrons are only coherent at ultralow temperatures in micron-sized samples. For classical waves in static samples, the wave is not inelastically scattered by the sample so that the wave remains temporally coherent throughout the sample even as its phase is random in space. In contrast, mesoscopic features of transport arise in disordered electronic systems only in samples with dimensions of several microns at ultralow temperatures. Mesoscopic electronic samples are intermediate in size between the microscopic atomic scale and the macroscopic scale. Electrons are typically multiply scattered within conducting samples so their dimensions are larger than the electron mean free path, which is on the scale of or larger than the microscopic atomic spacing and electron wavelength. At the same time, electronic samples are typically smaller than the macroscopic scale on which the wave function is no longer coherent. In contrast, monochromatic classical waves are generally temporally coherent over the average dwell time of the wave within large samples. It is therefore possible to explore the statistics of mesoscopic phenomena with classical waves. Such studies may also be instructive regarding the statistics of transport in electronic mesoscopic samples. Measurements can also be made in both the frequency and time domains. The impact of weak localization can be investigated in the time domain by measuring transmission following an excitation pulse or by Fourier transforming spectra of field multiplied by the spectrum of the exciting pulse. 

The connection between electronic and classical transport emerges as well from the equivalence proposed by Landauer \cite{54,55} of the dimensionless conductance {\textsl g} and the transmittance $T$, known as the``optical" conductance. The transmittance is the sum over all incident and outgoing channels of the transmission coefficient of flux. The phase of electrons arriving from a reservoir in different channels is randomized over the time of the measurement and the conductance is related to the incoherent sum of transmission coefficients over all channels, $\textsl{g}=\langle T\rangle=\langle \sum_{a,b=1}^N T_{ba}\rangle=\langle \sum_{a=1}^N T_a\rangle$. Measurements have been made of the statistics of transmission coefficients of the field, $t_{ba}$, intensity, $I_{ba}=|t_{ba}|^2$, and total transmission, $T_a=\sum_{b=1}^N T_{ba}$, for a single incident channel, $a$ and for the transmittance, $T$.

The experimental setup for measurements of microwave transmission in the Q1D geometry described in this chapter is shown in Fig. 1. 
\begin{figure}[htc]
\centering
\includegraphics[width=3in]{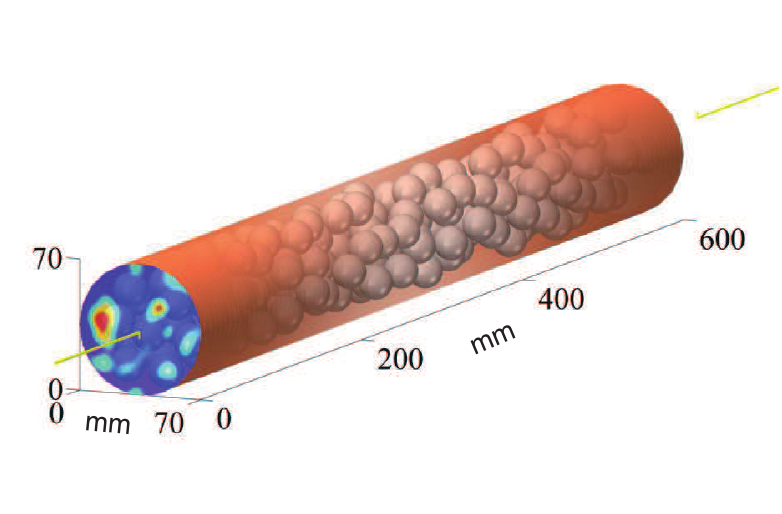}
\caption{ Copper sample tube containing a random medium with microwave source and detector antennas. Intensity speckle pattern is produced with a single source location. }\label{Fig1}
\end{figure}
Measurements are carried out in ensembles of random samples contained in a copper tube. The samples are random mixtures of alumina spheres with diameter of 0.95 cm and index of refraction of $n=3.14$ at a volume fraction of 0.068. Source and detector antennas may be translated over a square grid of points covering the incident and output surfaces of the sample and rotated between two perpendicular orientations in the planes of the sample boundaries. Spectra of the field transmission coefficient polarized along the length of a short antenna are obtained from the measurement of the in- and out-of-phase components of the field with use of a vector network analyzer. The intensity for a single polarization of the wave is the sum of the squares of the in- and out-of-phase components of the field. The sum of intensity across the output face for two perpendicular orientations of the detector antenna gives the total transmission. The field speckle pattern for each antenna position on the sample input is obtained by translating the detection antenna over the output surface. An example of an intensity speckle pattern formed in transmission is shown at the output of the sample tube in Fig. 1. The tube is rotated and vibrated momentarily after measurements are completed for each sample configuration to create a new and stable arrangement of scattering elements. In this way, measurements are made over a random ensemble of realizations of the sample. Field spectra can be Fourier transformed to yield the temporal response to pulsed excitation.  

Spectra of intensity, total transmission and transmittance normalized by the ensemble average values $s_{ba}=T_{ba}/\langle T_{ba}\rangle$, $s_a=T_a/\langle T_a\rangle$, and $s=T/\langle T\rangle$ in a single random configuration in two different frequency ranges are shown in Fig. 2. Fluctuations of relative intensity are noticeably suppressed in the higher frequency range as the degree of spatial averaging increases. For the ensemble represented in Fig. 2, var($s_a$) = 0.13 in the high frequency range and 3.88 in the low frequency range. Since waves are localized for var$(s_a)>2/3$  \cite{36}, this indicates that the wave is localized in the low frequency range.

\begin{figure}[htc]
\centering
\includegraphics[width=3in]{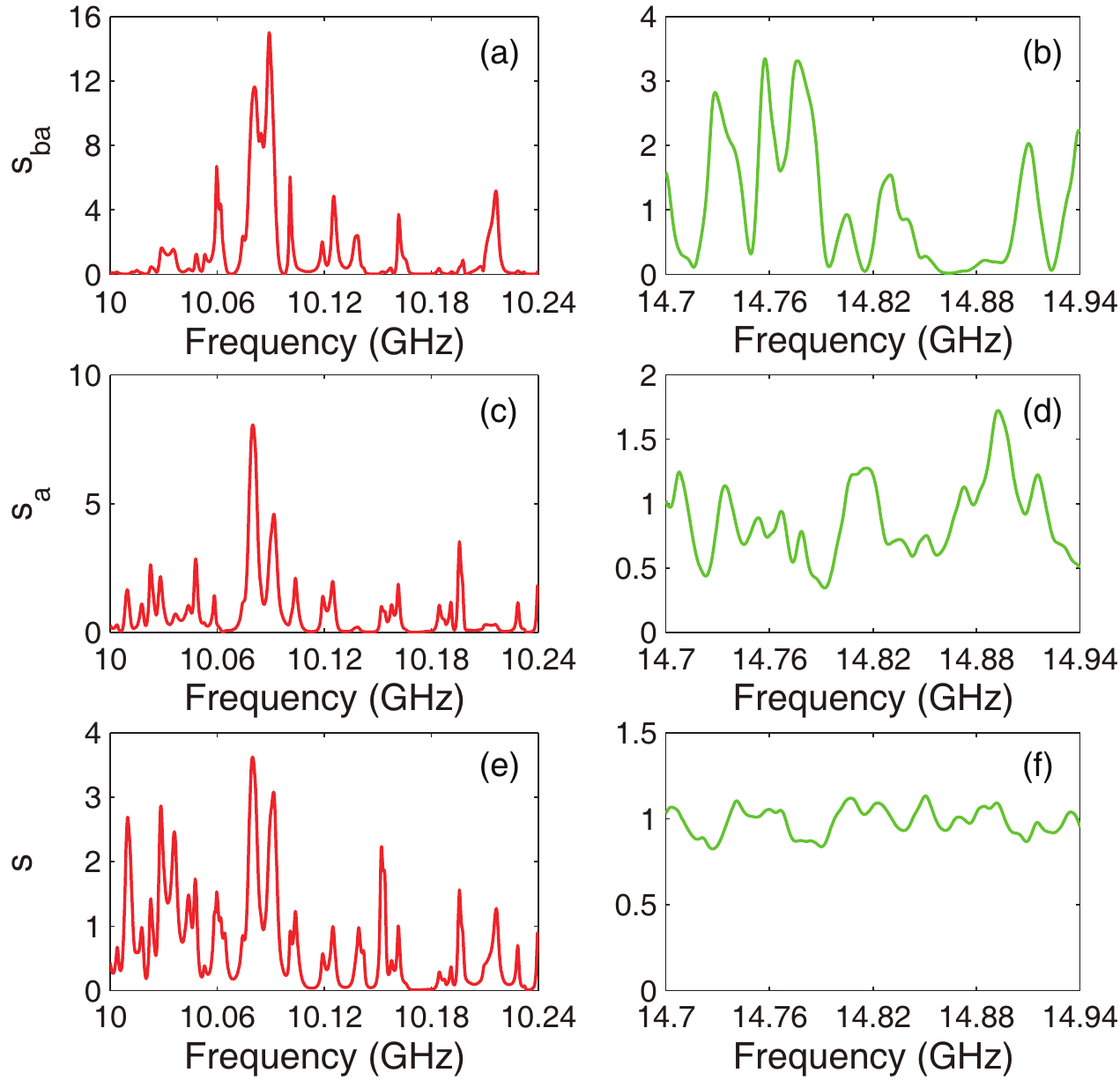}
\caption{Spectra of transmitted microwave intensity, total transmission, and transmittance relative to the ensemble average value for each in a single random configuration. The wave is localized in (a), (c),and (e) and diffusive in (b), (d),and (f). }\label{Fig2}
\end{figure}

It is instructive to consider the spectra in Fig. 2 from both the mode and channel perspectives. When the wave is localized, distinct peaks appear when the incident radiation is on resonance with a mode. The resonance condition holds for all source and detector positions and therefore sharp peaks remain even when transmission is integrated over space. When the wave is diffusive, many modes contribute to transmission at all frequencies and for all source and detector positions. The relative coupling strengths of a single polarization component of the intensity into and out of each of these modes has a negative exponential distribution and phases of the field transmission coefficient are random so that relative fluctuations will be suppressed with increased spatial averaging. From the channel perspective, many orthogonal transmission channels contribute to transmission for diffusive waves and the coupling to channels varies with source and detector positions. Fluctuations in the incoherent sum of this random jumble of orthogonal eigenchannels are therefore suppressed upon averaging over space. This suppresses the variance of transmission by a degree related to the number of channels that contribute substantially to transmission. This may be expressed quantitatively in term of the participation number of eigenvalue of the transmission matrix, $M\equiv (\sum_{n=1}^N \tau_n)^2/\sum_{n=1}^N \tau_n^2$. For diffusive waves, var($s_a$)$\sim 1/M$ and relative fluctuations are enhanced since the number of effective channels $M$ is smaller than the number of independent channels $N$. We will see below that for diffusive waves the spectrum of transmission eigenvalues is rigid so that the number of transmission eigenvalues above $1/e$ fluctuates by approximately unity and fluctuations of conductance $T$ are of order unity \cite{1, 4, 56, 70, 70a, 70b}. This results in universal conductance fluctuations which are independent of the sample size for Q1D sample \cite{1, 4}.

The localization transition may be charted in terms of a variety of related localization parameters, all of which can be measured for classical waves. In addition to $\delta$ and the average over a random collection of samples of the dimensionless conductance, ${\textsl g}=\langle T\rangle$, measurements of fractional fluctuations of intensity or total transmission characterize the nature of the wave in random systems. In the diffusive limit, the variance of total transmission relative to the average value of total transmission over a random ensemble of statistically equivalent samples is inversely proportional to {\textsl g}, var$(s_a)=2/3{\textsl g}$ \cite{3,30,31,32,36}. Since the wave is localized for ${\textsl g}<1$, localization occurs when var$(s_a)>2/3$. Perhaps the most easily accessible experimental localization parameter is the variance of fractional intensity, which can be expressed as var$(s_{ab})=$ 1 + 4/3{\textsl g} \cite{31}. The localization threshold at {\textsl g} = 1 corresponds to var($s_{ab}$)=7/3. var($s_a$) and var($s_{ab}$) remains useful localization parameters even for localized waves. Fluctuations are relatively insensitive to absorption as compared to measurements of absolute transmission \cite{36}. Mesoscopic fluctuations are directly tied to intensity correlation within the sample \cite{20,21,22,24,72,72a}. The fractional correlation of intensity at two points on the output surface or between two transmission channels, {\it b} and $b^\prime$, is equal to the variance of relative total transmission, $\kappa=\langle \delta s_{ba} \delta s_{b^\prime a}\rangle=$var($s_a$). It is equal to $\langle M^{-1}\rangle$ in the diffusive limit, which is enhanced over the value of $1/N$ that would be expected if mesoscopic correlation were not present.

The relationships between key localization parameters mentioned above arise since the nature of propagation in disordered Q1D samples, in which the wave is thoroughly mixed in the transverse directions, depends only on a single dimensionless parameter \cite{8}. For diffusive waves, $\delta={\textsl g}=2/3$var$(s_a)=2/3\kappa=2/3\langle M^{-1} \rangle$. The relationships var$(s_a)=\kappa$ and var($s_{ab}$)=2var($s_a$)+1 hold through the localization transition, but the relationship between the other variable does not. However, we anticipate these relationships will change in a manner that can be described in terms of a single parameter. Other classical wave measurements that indicate the closeness to the localization threshold are coherent backscattering \cite{17,18,19,29} and the transverse spread of intensity in steady state or in the time domain \cite{65,66,67,68,73}. The width of the coherent backscattering peak gives the transverse spread of the wave on the incident surface and hence the transport mean free path $\ell$, from which the value of $k\ell$ can be found.

\section{Modes}
We find that the fields at any point in the sample may be expressed as a superposition of the field associated with the excitation of all the modes in the sample. This superposition is a sum of products for each mode of the $j$ polarization component of the spatial variation of the mode, $a_{n,j}({\bf r})$, and the frequency variation of the mode, which depends only upon the central frequency of the mode $\omega_n$ and its linewidth, $\Gamma_n$,
\begin{equation}
E_j({\bf r,\omega})=\sum_{n}a_{n,j}({\bf r})\frac{\Gamma_n/2}{\Gamma_n/2+i(\omega -\omega_n)}=\sum_{n}{a_{n,j}({\bf r})\varphi_n(\omega)}.
\end{equation}
The frequency variation of the $n^{th}$ mode $\varphi_n(\omega)$ is given by the Fourier transform of $\exp (-\Gamma_n t/2)\cos\omega_n t$ for $t > 0$. Equation (1) can be fit simultaneously to the field at a large number of points on the output speckle patterns in a single configuration since all spectra share a common set of $\omega_n$ and $\Gamma_n$. Armed with the values of $\omega_n$ and $\Gamma_n$, we find $a_{n,j}({\bf r})$ and hence the speckle pattern for each of the modes. 

The transmission spectrum is determined by the variation with position of the field amplitudes $|a_{n,j}({\bf r})|$ and phases over the transmitted speckle patterns for the modes. The contribution of individual modes to transmission can be seen in the spectrum of total transmission near the single strong peak at 10.15 GHz shown in Fig. 3(a) in a random sample of length $L=61$ cm. The asymmetrical shape for the line in both intensity and total transmission indicates that more than a single mode contributes to the peak. The modal analysis of the field spectra shows that three modes contribute substantially to transmission over this frequency range. Spectra of the total transmission for the three modes closest to 10.15 GHz acting independently are plotted in Fig. 3(a). The integrated transmission for the 28th and 29th mode found in the spectrum starting at 10 GHz are each greater than for the measured peak indicating that these modes interfere destructively. The intensity and phase patterns for these two modes are shown in Figs. 3b-e. Aside from a difference in the average value of transmission, the intensity speckle patterns of the two modes are nearly the same. The distributions of phase shift at 10.15 GHz for the two modes are also similar except for a constant phase difference between them of $\Delta \varphi=1.02\pi$ rad. The similarity between the speckle patterns for these overlapping modes suggests that these modes are formed from coupled resonances within the sample which overlap spatially and spectrally. We expect that such resonances peaked at different locations will hybridize to form modes of the system. Such modes may be close to symmetric and antisymmetric combinations of the two local resonances. This would produce similar intensity speckle patterns at the output with a phase shift of $\sim \pi$ rad between the modes. The similarity in the intensity speckle patterns of these adjacent modes and the uniformity of the phase shift across the patterns of these modes allows for interference between modes across the entire speckle pattern. The similarity between modes is most evident in a sample configuration such that a pair of nearest neighbor modes are particularly close in frequency. This is a point of anticrossing which arises because of level repulsion inside the sample \cite{62,74}. The magnitude of the field inside a 1D sample is seen to be the same throughout the sample. Modes are orthogonal by virtue of a change of phase of $\pi$ rad along the length of the sample. At the anticrossing, the fields at the sample output for the two modes are the same except for a change in phase close to $\pi$ rad. The Thouless number, which equals the dimensionless conductance for diffusive waves, provides a key measure of the dependence of transmission on the underlying characteristics of modes. The modal decomposition of transmission spectra for the ensemble from which the configuration is analyzed in Fig. 3 is drawn gives $\delta$ = 0.17 \cite{51}.
\begin{figure}[htc]
\centering
\includegraphics[width=3.5in]{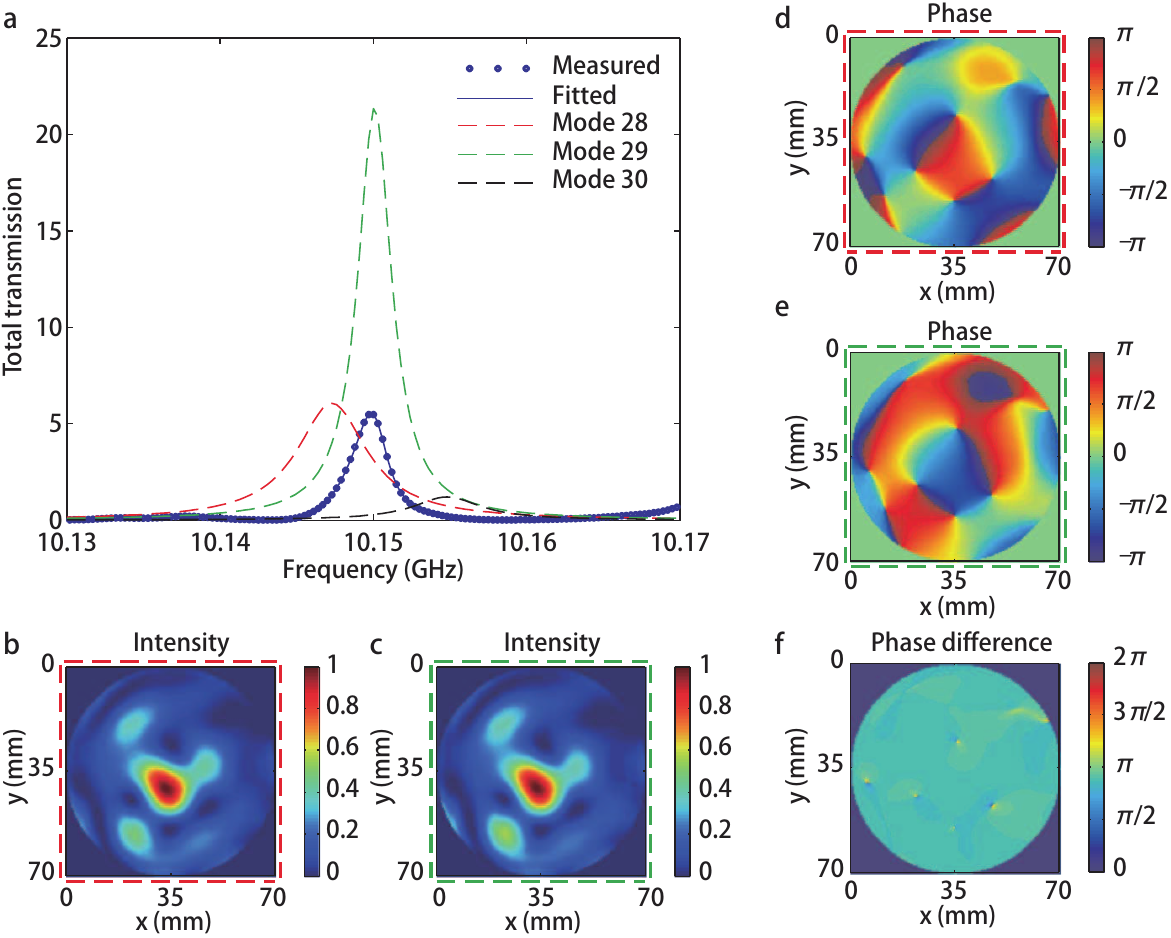}
\caption{(a) Three modes contribute to the asymmetric peak in the total transmission spectrum. Modes 28, 29 and 30 are in order of increasing frequency. Intensity speckle patterns for modes 28 and 29 are shown in (b) and (c) and the corresponding phase patterns are shown in (d) and (e). (f) The phase in mode 29 is shifted by nearly a constant of 1.02$\pi$ rad relative to mode 28. (Ref. \cite{51})}\label{Fig3}
\end{figure}

The statistics of level spacing was first considered by Wigner \cite{76} in the context of nuclear levels probed in neutron scattering. He conjectured the eigenvalues of the Hamiltonian matrix would have the same statistics as the spacing of eigenvalues of a large random matrix with Gaussian elements. Agreement was found between the spacing between peaks in the scattering cross section and Wigner's surmise for the spacing of eigenvalues of random Hamiltonian matrices. However, the analysis of spectra of nuclear scattering cross sections was done in samples with relatively sharp spectral lines. We have seen above that even when $\delta<1$, a number of lines may coalesce into a single peak. A comparison of level spacing statistics in samples with different values of modal overlap $\delta$ in which the phase of the scattered wave can be measured is therefore of interest in forming a picture of the statistics of transmission. Progressively stronger deviations from the Wigner surmise are found for decreasing values of $\delta$.

\begin{figure}[htc]
\centering
\includegraphics[width=3.5in]{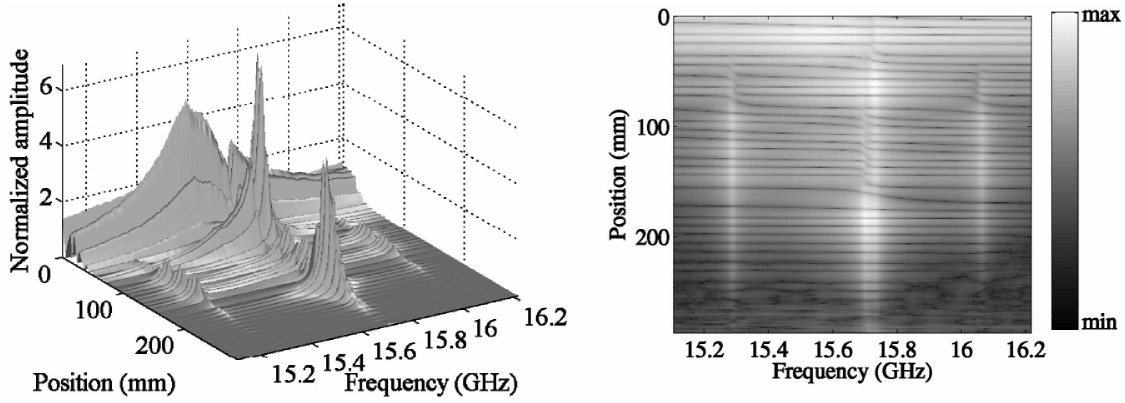}
\caption{(a) Spectra of the field amplitude at each point along a random sample with spectrally overlapping peaks normalized to the amplitude of the incident field. (b) Top view of (a) in logarithmic presentation. (Ref. \cite{61})}\label{Fig4}
\end{figure} 
It has not been possible to access the field distribution within the interior of multiply scattering three-dimensional samples, but spatial distributes can be examined in one- and two-dimensional samples \cite{59,61,62}. The presence of both isolated and overlapping modes within the same frequency range has been observed in measurements of field spectra along the length of slotted single-mode random waveguides. The waveguides contained randomly positioned binary dielectric elements and a smaller number of low index Styrofoam elements. Measurements were carried out in the frequency range of a pseudogap associated with the first stop band of a periodic structure of consecutive binary elements. The density of states is particularly low in the frequency range of the band gap so that $\delta<1$. When spectrally isolated lines are found, they are strongly peaked in space and their intensity spectra at each point in the sample is Lorentzian with the same width at all points within the sample. When modes overlap spectrally, however, spectral peaks have complex shapes which vary with position within the sample and the spatial intensity distribution is multiply peaked. Mott \cite{77} argued that interactions between closely spaced levels in some range of energy in which $\delta<1$ would be associated with two or more centers of localization within the sample. Pendry \cite{78} showed that the occasional overlap of electronic states would dominate transport since regions in which the value of the electron wave function is high would not be far from both the input and output boundaries. Since the wave can then find a ready path through the sample, such modes are relatively short-lived and spectrally broad. This enhances the contribution of coupled resonances to transport. Such multiply peaked and spectrally overlapping excitations within the sample, termed``necklace states" by Pendry, are important in transmission since they arise in the localized regime and transport through isolated modes is typically small and over a narrow linewidth.

The variation in space and frequency of the amplitude of the waves within the pseudogap in a single random configuration is shown in Fig. 4. An additional ripple is observed in Fig. 4(b) in the intensity variation through the sample, corresponding to a phase shift of $\pi$ rad, each time the frequency is tuned though a mode. The decomposition of field spectra inside the waveguide within the pseudogap into the modes and a background which varies slowly in frequency is shown in Fig. 5. The slowly varying background shown in Fig. 5(a) is the fit of a polynomial in the difference in frequency from a point in the middle of the spectral range considered. This background is presumably related to off-resonance excitation of many modes on either side of the band gap. The mode structure within the single-mode waveguide sample changes when a spacing is introduced between two parts of the sample and is increased gradually. A succession of mode hybridizations is observed with increasing spacing as a single mode tends to shift in frequency until it encounters the next mode. As the spacing is increased, the mode that had been moving becomes stationary and the next mode begins to move \cite{74}. Simulations have shown that changing the index of a single scatterer leads to mode hybridization in 2D random systems, in which a single peak may be transformed to multiple peaks \cite{62}.  
\begin{figure}[htc]
\centering
\includegraphics[width=2.5in]{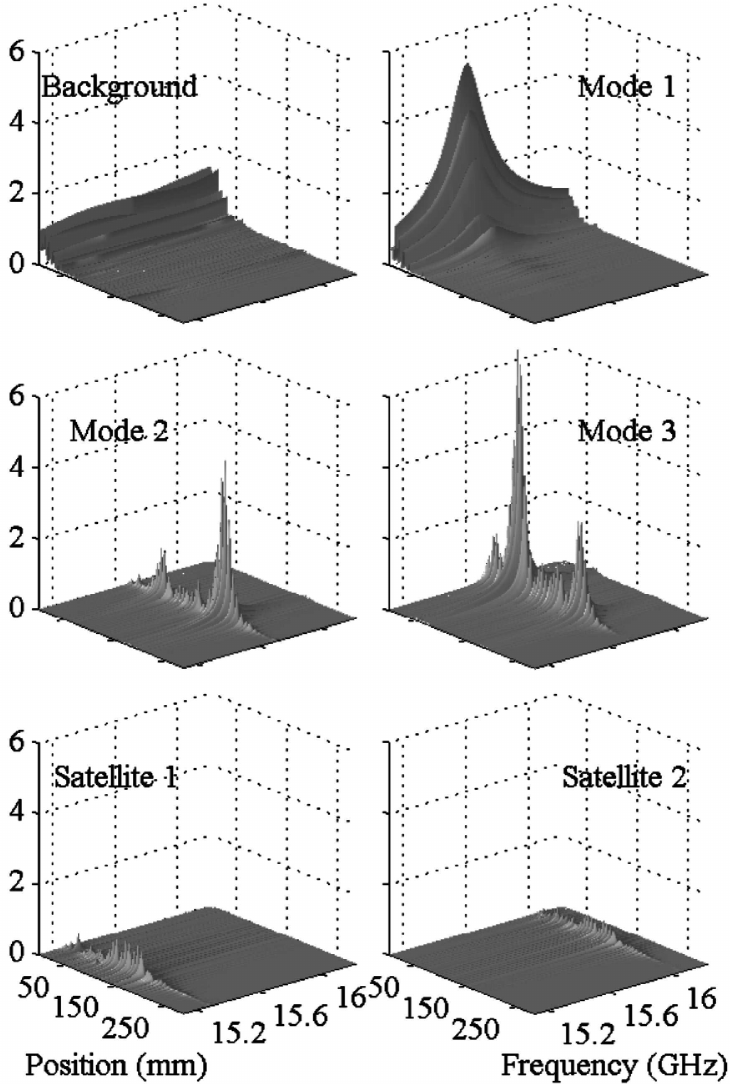}
\caption{Decomposition of field pattern in Fig. 4 into a slowly varying polynomial term and five modes. (Ref. \cite{61})}\label{Fig5}
\end{figure}

One-dimensional localization has also been observed in optical measurements in single-mode optical fibers \cite{60} and in single-mode channels that guide light within photonic crystals \cite{79,79a,79b}. When the structure bracketing the channel is periodic, the velocity of the wave propagating down the channel experiences a periodic modulation so that a stop band is created. When disorder is introduced into the lattice, modes with spatially varying amplitude along the channel are created. Modes near the edge of the band gap are long lived and readily localized by disorder. An example of spectra of vertically scattered light versus frequency for light launched down a channel through a tapered optical fiber is shown in Fig. 6. The inset shows the disordered sample of holes with random departure from circularity in silicon-on-insulator substrates at a hole filling fraction of $f\sim0.30$. 
\begin{figure}[htc]
\centering
\includegraphics[width=2.5in]{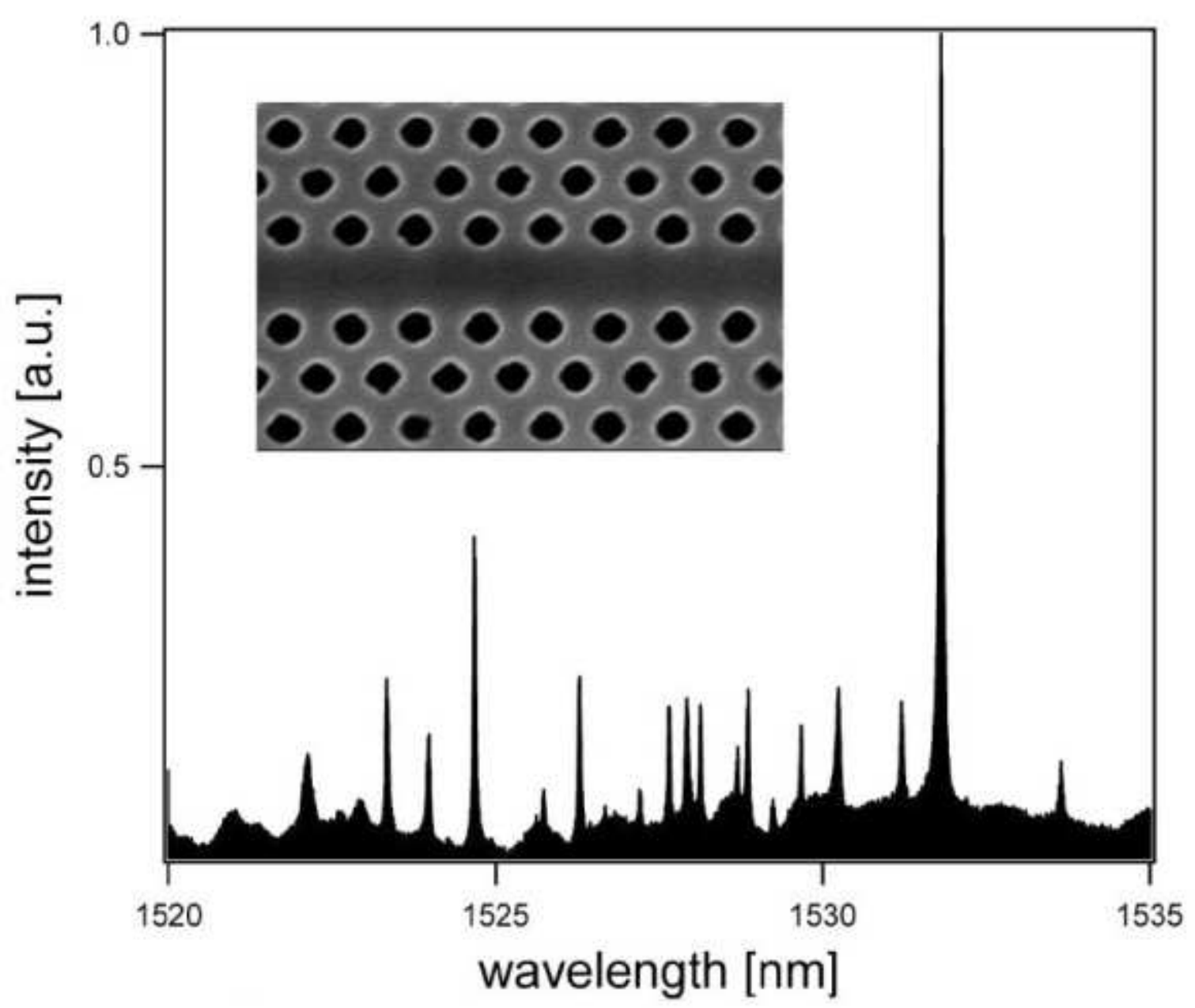}
\caption{Spectrum of wave transmitted to a region within a single-mode photonic crystal waveguide near the short wavelength edge of the first stop band at $\sim1520$ nm. The channel surrounded by irregular holes is shown in the inset. (Ref. \cite{79})}\label{Fig6}
\end{figure}

The modal decomposition method described above can be applied to localized waves for which modal overlap is relatively small. The impact of modes can be seen in the changing decay rate of transmission following pulsed excitation, even for diffusive waves. The slowing down of the decay rate with time will become more pronounced in samples in which the wave is more strongly localized \cite{81,82,83}. In the diffusive limit, the transverse extent of the modes is large and the wave is coupled to its surroundings through a large number of speckle spots. One expects therefore that the decay rate of all modes will approach the decay rate of the lowest diffusion mode \cite{84,85}, $1/\tau_1=\pi^2D/(L+2z_0)^2$,	after a time $\tau_1$ in which higher order modes with decay rates $1/\tau_n=n^2\pi^2D/(L+2z_0)^2$ have largely decayed. Here, $n$ is the order of the diffusion mode and $z_0$ is the length beyond the boundary of the sample at which the intensity inside the sample extrapolates to zero. We find that pulsed transmission deviates increasingly from the diffusion model in nominally diffusive samples with ${\textsl g}>1$ as the value of {\textsl g} decreases and the measured value of $\kappa$ increases. Strong deviations from pulsed transmission in diffusing samples far from the localization threshold are observed for microwave radiation, light and ultrasound \cite{80,73,68}. 
\begin{figure}[htc]
\centering
\includegraphics[width=4in]{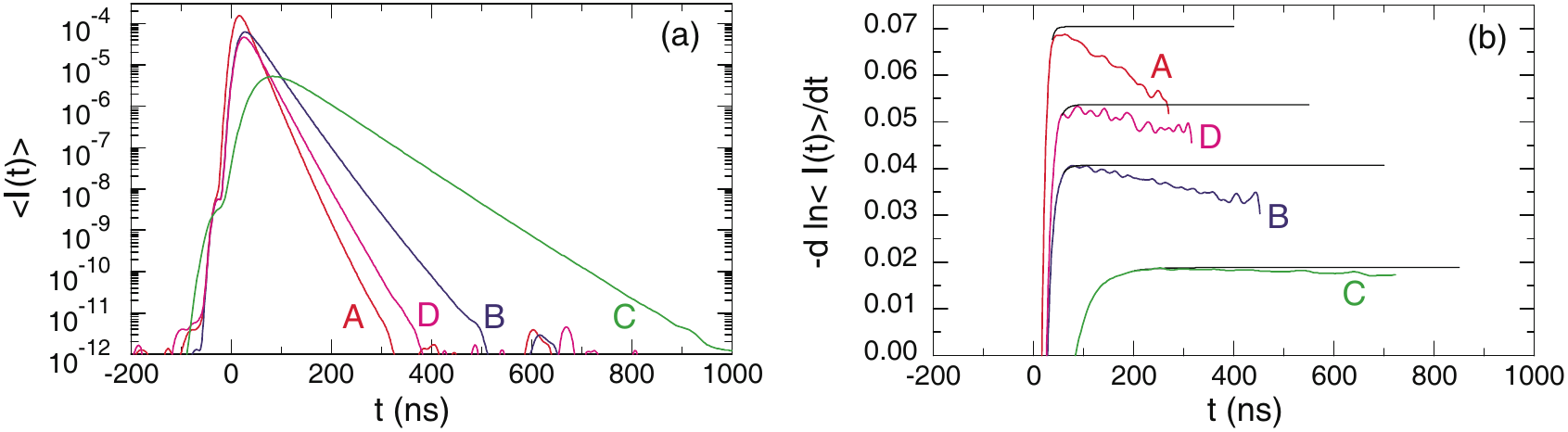}
\caption{(a) Average pulsed transmitted intensity in samples of alumina spheres with lengths, $L = 61$ cm (A), 90 cm (B and D), and 183 cm (C). Sample D is the same as sample B except for the insertion of a titanium foil inserted along the length of the sample tube D to increase absorption. (b) Temporal derivative of the intensity logarithm gives the rate $\gamma$ of the intensity decay due to both leakage out of the sample and absorption. (Ref. \cite{80})}\label{Fig7}
\end{figure}

Measurements of pulsed transmission through a random sample of alumina spheres at low-density in samples of different length and absorption with values of $\kappa=$ 0.09, 0.13, 0.25, and 0.125 are shown in Fig. 7 in samples A-D, respectively \cite{80}. The decay rate of intensity is seen to deviate from the constant rate of the diffusive limit and is seen in Fig. 7(b) to decreases at a nearly constant rate. A linear falloff of the decay rate would be associated with a Gaussian distribution of decay rates for the modes of the medium \cite{81}. A slightly more rapid decrease of the decay rate is associated with a slower than Gaussian falloff of the distribution of mode decay rates. The slowing down of the decay rate at long times reflects the survival of more slowly decaying modes \cite{80,81}. The distribution of modal decay rates is related to the Laplace transform of the transmitted pulse intensity.

Sample D is the same as sample B except for the increased absorption due to a titanium foil inserted along the length of the sample tube. The variation with time of the decay rate in sample D is the same as that in sample B except for an additional constant decay rate in sample D due to absorption. This shows that, at the low level of absorption in these samples, scattering rates are not affected by absorption and that the effect of absorption simply introduces a multiplicative exponential decay, which is the same for all trajectories at a given time. Thus the degree of renormalization of transport due to weak localization involving the interference of waves following time-reversed trajectories that return to a point in the medium is not affected by absorption. We note that the fractional reduction of the decay rate is greater at a given time delay in shorter samples with higher values of {\textsl g}. This is because the length of trajectories of partial waves within the medium is the same for all samples at a given delay but the number of crossings of trajectories is greater when the paths are confined within a smaller volume. 

The temporal variation of transmission can also be described in terms of the growing impact of weak localization on the dynamic behavior of waves, which can be expressed via the renormalization of a time-dependent diffusion constant or mean free path \cite{86}. The decreasing decay rate has also been explained using a self-consistent local diffusion theory for localization in open media \cite{87,88}. The theory uses a one loop self-consistent calculation of an effective diffusion coefficient that falls with increasing depth inside the sample. The spatial variation of the local diffusion coefficient reflects the increasing fraction of returns of wave trajectories to a point with greater depth due to the lengthened dwell time in the sample. This theory gives excellent agreement with recent measurements of non-diffusive decay of pulsed ultrasound transmission through a sample of aluminum spheres as seen in Fig. 8 \cite{68}.
\begin{figure}[htc]
\centering
\includegraphics[width=2.5in]{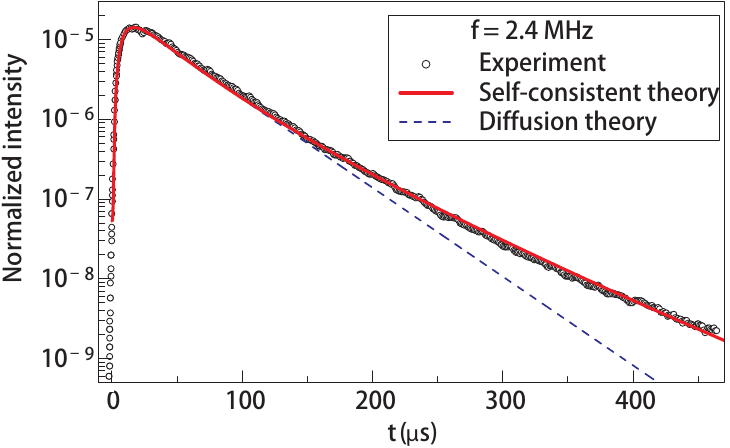}
\caption{Averaged time-dependent transmitted intensity $I(t)$ normalized so that the peak of the input pulse is unity and centered on $t=0$, at representative frequencies in the localized regimes. The data are fit by the self-consistent theory (solid curve). For comparison, the dashed line shows the long-time behavior predicted by diffusion theory. (Ref. \cite{68})}\label{Fig8}
\end{figure}
The value of {\textsl g} is just beyond the localization threshold as determined from measurements of var($s_{ba}$), which is slightly above the value of 7/3 predicted as the localization threshold. The intensity distribution on the output face of the sample is found to be multifractal as predicted near the threshold for Anderson localization \cite{81,89}. However, the value of $k\ell$ in this sample is $\sim1.5$, which would indicate the wave is diffusive. Measurements of the spread of intensity on the sample output with increasing time delay show a trend towards an exponential decay of intensity on the output plane at later times supporting the localization of the wave. A similar approach to an exponential decay of intensity with transverse displacement on the sample output from the point of injection of the pulse on the incident surface has been observed by Sperling {\it et al.} \cite{73} in optical measurements through a slab of titania particles. When $k\ell<4.5$, the variance of the spatial intensity distribution reaches a peak value and then actually falls. This is taken as support of localization of the wave at $k\ell>1$. However, this criterion for localization is directly tied to the Thouless criterion for localization $\delta=1$. It might also be that at later times, longer lived modes, which are more confined in space are more heavily represented and dominate the spatial distribution. Though each of these modes will not spread in time, the modes that survive with increasing time delay would be the more strongly confined modes and would lead to a falling variance of the spatial intensity distribution in time. Such states may be “prelocalized” with a slower falloff in space than exponential but still faster than for diffusive waves \cite{90,91}.

The slowing of the spread of the transmitted wave in the transverse direction can also be seen in transverse localization in samples which are uniform in the longitudinal direction. This has been observed in a 2D periodic hexagonal lattice with superimposed random fluctuations \cite{66}. The structured sample is created by first illuminating the photorefractive sample with a hexagonal optical pattern and then with a random speckle pattern of varying strength. A transition from a diffusive to a localized wave in the transverse plane is seen in the output plane with the ensemble average of the spatial intensity distribution changing from a Gaussian to an exponential function centered on the input beam as the thickness of the sample increases. Since the wave incident upon the sample, which is uniform along its length, is paraxial, it is not scattered in the longitudinal direction and travel time through the sample is proportional to the sample thickness. Transverse localization is also observed in an array of disordered waveguides lattices \cite{67}.

Measurements of pulsed microwave transmission of more deeply localized waves transmitted through a Q1D sample of random alumina spheres of thickness approximately 2.5 times the localization length, are shown in Fig. 9 \cite{92}. 
\begin{figure}[htc]
\centering
\includegraphics[width=4in]{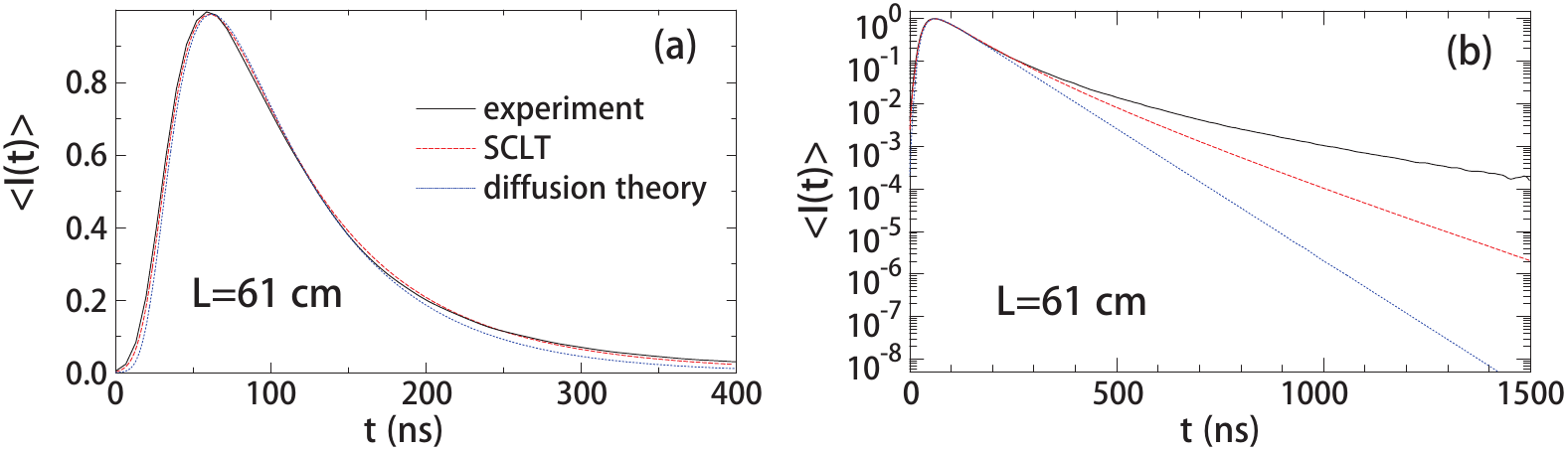}
\caption{(a) Fit of self-consistent localization theory (SCLT) (dashed curve) at early times to the average intensity response (solid curve) to a Gaussian pulse with $\sigma_\nu=15$ MHz in a sample of length $L=61$ cm and the result of classical diffusion theory (dotted curve); (b) semilogarithmic plot of $\langle I(t) \rangle$ reaching to longer times; In (a) and (b), the curves are normalized to the peak value. (Ref. \cite{92})}\label{Fig9}
\end{figure}
The impact of absorption was removed statistically by multiplying the average measured intensity distribution by $\exp(t/\tau_a)$ \cite{36,93}. For times near the peak of the transmitted pulse, diffusion theory corresponds well with the measurements of $\langle I(t)\rangle$. For times up to 4 times $\tau_D=L^2/\pi^2D$, the decay rate of the lowest diffusion mode, pulsed transmission is in accord with self-consistent localization theory, but transmission decays more slowly for longer times. This indicates the inability of this modified diffusion theory to capture the decay of long-lived localized states. Such states are included in a position-dependent diffusion theory \cite{94} that is in accord with simulations of the steady-state intensity profile within random systems \cite{95}. The difference between self-consistent localization theory and the theory for position-dependent diffusion is seen to be precisely in the ability of the latter to include the impact of long-lived resonant states \cite{94,95}.

Destructive interference between neighboring modes together with the distribution of mode transmission strengths and decay rates can explain the dynamics of transmission. The average temporal variation of total transmission due to an incident Gaussian pulse is found from the Fourier transform of the product of the field spectrum and the Gaussian pulse. The progressive suppression of transmission in time by absorption may be removed by multiplying $\langle T_a(t)\rangle$ by $\exp(t/\tau_a)$ to give, $\langle T_a^0(t)\rangle=\langle T_a(t)\rangle \exp(t/\tau_a)$ \cite{36,93}. With the influence of absorption upon average transmission removed, decay is due solely to leakage from the sample. The measured pulsed transmission corrected for absorption is shown as the solid curve in Fig. 10 and is compared to the incoherent sum of transmission for all modes in the random ensemble corrected for absorption, $\sum_{n}{T_{an}^0(t)}$, shown as the dashed curve in the Fig. 10. $\sum_{n}{T_{an}^0(t)}$ is substantially larger than $\langle T_a^0(t)\rangle$ at early times, but converges to $\langle T_a^0(t)\rangle$ soon after the peak. Though transmission associated with individual modes rises with the incident pulse, transmission at early times is strongly suppressed by the destructive interference of modes with strongly correlated field speckle patterns such as those shown in Fig. 3. At later times, random frequency differences between modes leads to additional random phasing between modes and averaged pulsed transmission approaches the incoherent sum of decaying modes. The decay of $\langle T_a^0(t)\rangle$, shown as the solid curve in Fig. 10, is seen to slow considerably with time delay reflecting a broad range of modal decay rates. 
\begin{figure}[htc]
\centering
\includegraphics[width=2.5in]{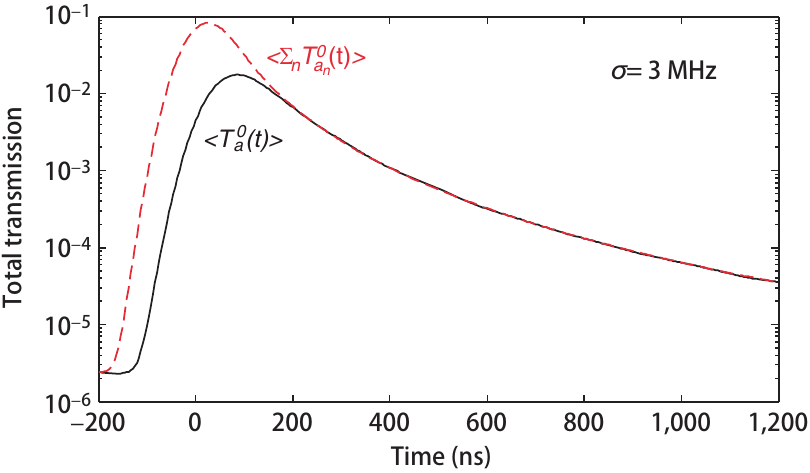}
\caption{Semilogarithmic plot of the ensemble average of pulsed transmission and the incoherent sum of transmission due to all modes in the random ensemble. (Ref. \cite{51})}\label{Fig10}
\end{figure}

\begin{figure}[htc]
\centering
\includegraphics[width=2.5in]{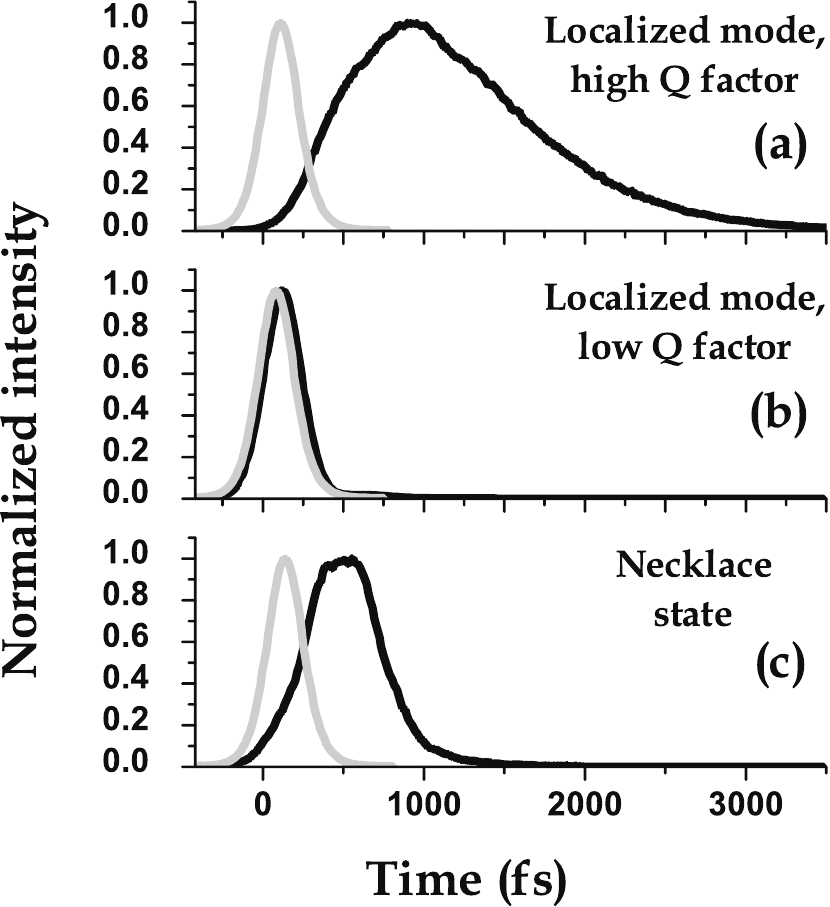}
\caption{Time-resolved transmission data. In (a) and (b) the sample is excited on resonance with a sharp transmission peak, with small and large linewidths. In (c) a nearly symmetric pulse shape is observed that exhibits a fast decay time and a relatively large delay, typical for multiple-resonance necklace states. Sample thickness: 250 layers. Gray curves: instrumental response (cross correlation between probe and gate), corrected for the delay introduced by the effective refractive index of the sample. (Ref. \cite{63})}\label{Fig11}
\end{figure}
Measurements by Bertolotti {\it et al.} \cite{63} of pulsed infrared transmission through random layers of porous silicon with different porosity produced by controlled electrochemical etching of silicon show that the pulse profiles depend on the degree of spectral overlap of excited modes. As the number of layers increases, spectra become sharper since propagation of a paraxial beam in the structure is essentially one dimensional and $\delta$ falls with sample thickness. When the pulse excites an isolated resonance peak, the decay rate of the falling edge of the transmitted pulse is seen in Figs. 11(a) and (b) to be larger for the narrower spectral peak. The delayed rise in transmission seen in Fig. 11(a) suggests, however, that more than a single mode is involved since the interaction of a pulse with a single mode would lead to a prompt rise in transmission as the pulse enters the sample, after which intensity decays at a constant rate. When the spectrum of the exciting pulse overlaps several spectral lines, a symmetrical profile for the transmitted pulse is observed, as seen in Fig. 11(c). These spectrally overlapping states form necklace states with a series of intensity peaks along the sample which provide a path for the wave through the medium. Such states would be expected to be short-lived. 

The relationship of pulsed transmission to the transmission spectrum of microwave radiation through a sample of random dielectric spheres with $\delta = 0.43$ is shown in Fig. 12 \cite{92}. The decay is slow when the spectrum of the pulse overlaps a single narrow mode and fast when two peaks fall within the spectrum of the incident pulse. In the latter case, the transmitted intensity is significantly modulated at the frequency difference between the modes.
\begin{figure}[htc]
\centering
\includegraphics[width=4in]{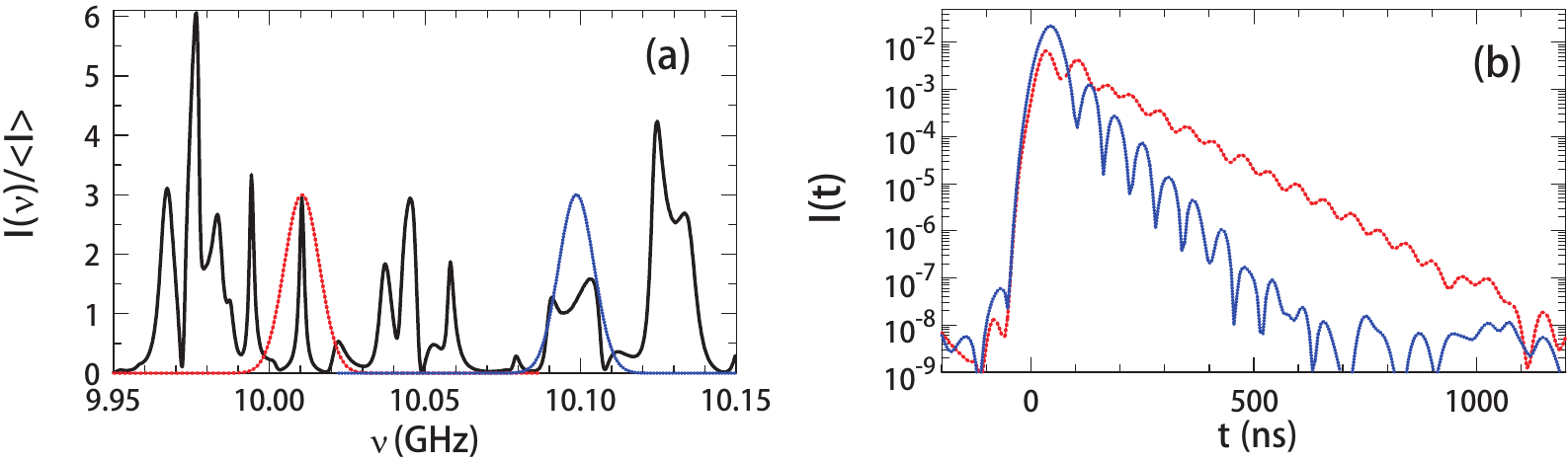}
\caption{(a) Transmitted intensity spectrum in a random sample of $L=40$ cm and Gaussian spectra of incident pulses peaked at the center of the isolated line and overlapping lines. (b) Intensity responses to the Gaussian incident pulses with spectral functions shown in (a). (Ref. \cite{92})}\label{Fig12}
\end{figure}

In addition to the reduction of the leakage rate with increasing delay observed in diffusive samples, the variance of relative intensity fluctuations and the degree of intensity correlation also increase with time delay \cite{96,97,98}. The field correlation function with displacement and polarization rotation in pulsed transmission is the same as in steady state \cite{96}. This reflects the Gaussian statistics within the speckle pattern of a given sample configuration and time delay. The intensity correlation function at a given time delay depends on the square of the field correlation function and the degree of intensity correlation in the same way as in steady state, but the degree of extended range correlation $k_\sigma(t)$ depends on delay time and on the spectral bandwidth of the pulse $\sigma$. The probability distribution functions of intensity at various delay times have the same form as for steady state propagation and depend upon a single parameter, which is the variance of the total transmission relative to its average over the ensemble, which equals the degree of intensity correlation at that time, var($s_a(t)$)=$\kappa_\sigma(t)$. The time variation of $\kappa_\sigma(t)$ reflects the number of modes and the degree of correlation in the speckle pattern of the modes. Since strong correlation in the speckle patterns of a number of modes tends to produce a single transmission channel formed from these modes while correlation at any time is directly related to the number of channels contributing to transmission at that time, modal speckle correlation tends to increase the degree of intensity correlation. 

For narrowband excitation, $\kappa_\sigma(t)$ first falls before increasing at later times since transmission at early times is dominated by a subset of short-lived modes among all the modes overlapping the spectrum of the pulse that promptly convey energy to the output \cite{98}. At later times, only the long-lived modes contribute to transmission and so the degree of correlation increases with time. But at intermediate times, when both short- and long-lived modes contribute to transmission, the number of modes and hence the number of channels contributing is relatively high. This is inversely proportional to the degree of correlation so that $\kappa_\sigma(t)$ reaches a minimum.
\begin{figure}[htc]
\centering
\includegraphics[width=2.5in]{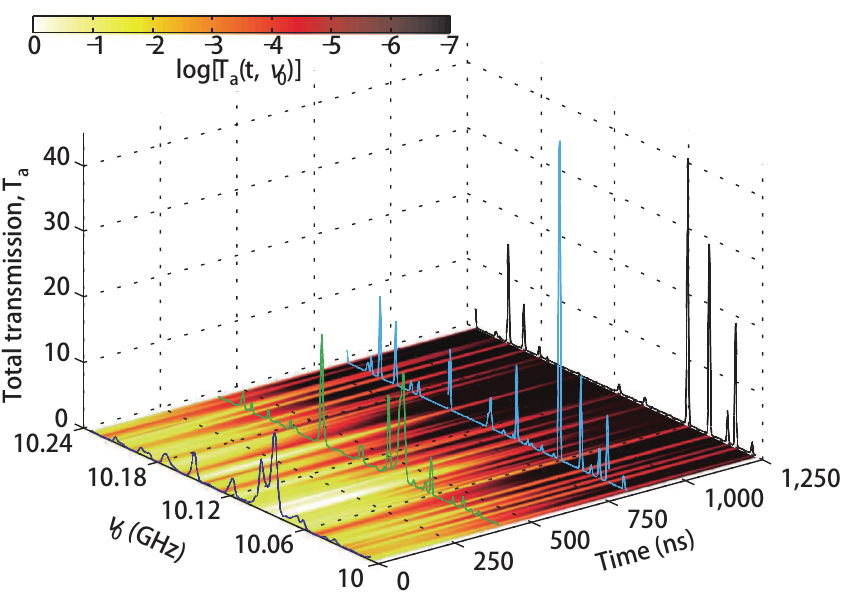}
\caption{Logarithm of time-frequency spectrogram of total transmission plotted in the x-y plane following the color bar. The central frequency $\nu_0$ of the incident Gaussian pulse of linewidth $\sigma=$50.85 MHz is scanned. Each of the four spectra of total transmission at different delay times are normalized to the total transmission at that time. (Ref. \cite{51})}\label{Fig13}
\end{figure}

The changing distribution of modes contributing to transmission is seen in the time-frequency spectrogram for a sample with $L=61$ cm and $\delta=0.17$ in Fig. 13. The spectrogram is formed from measurements of transmission at the output of a sample for an incident Gaussian pulse with width $\sigma$ of its Gaussian spectrum as the central frequency of the pulse is tuned. The decay rate of the peak intensity of the mode at long times when isolated modes emerge in the time-frequency spectrogram is equal to the linewidth of the mode $\Gamma_n$ to within experimental error of 10\%.

\section{Lasing in localized modes}
The nature of propagation divides according to the character of the modes of the medium. For $\delta>1$, transport can be described in terms of diffusing particles of the wave with transmission falling inversely with sample thickness, while for $\delta<1$ transport is via tunneling through localized or multipeaked modes with average transmission falling exponentially. Since the location and intensity of pump excitation within the sample and the lifetime of emitted photons within the gain region in which stimulated emission occurs depend upon the character of modes, lasing characteristics depends crucially upon the value of $\delta$. 

For $\delta \gg 1$ in random amplifying samples, nonresonant random lasing occurs. This can be described in terms of the densities of diffusing pump and emission photons and their coupling to energy levels whose occupation are described in terms of rate equations \cite{99,100,101}. Very different behavior arises in the regime $\delta \sim$1, which can arise in strongly scattering but still diffusive samples which are not more than a few wavelengths thick \cite{102,102a} and in 2D samples \cite{102b}, in which the laser beam is tightly focused to create a small excitation volume \cite{102X,102Y}. A small number of spectral peaks may then be observed in emission. These peaks sharpen up in the presence of gain due to enhanced stimulated emission in longer lived, spectrally narrow modes \cite{102}.”

Letokhov \cite{103} considered the lasing threshold in a spherical sample with uniform gain which is directly analogous to the critical condition for a nuclear chain reaction. Lasing occurs when on average more than one new photon is created for each photon that escapes the medium. Lasing was subsequently considered in granular media and in colloidal samples composed of dielectric particles in dye solution. Lasing in amplifying colloids reported by Lawandy {\it et al.} \cite{104} is of particular interest since the strength of scattering and amplification can be controlled independently. A narrowing of emission and a shortening of the emitted pulse was observed above a threshold in pump power. A comparison of the emission spectrum in a neat dye solution and in colloidal solutions is shown in Fig. 14. The original studies were carried out in weakly scattering samples excited over transverse dimensions much greater than the sample thickness, which itself was not much thicker than the mean free path. Wiersma {\it et al.} \cite{105} suggested that the observations reported could be due to scattering of light into transverse directions and its subsequent redirection out of the sample by another scattering event. 
\begin{figure}[htc]
\centering
\includegraphics[width=2in]{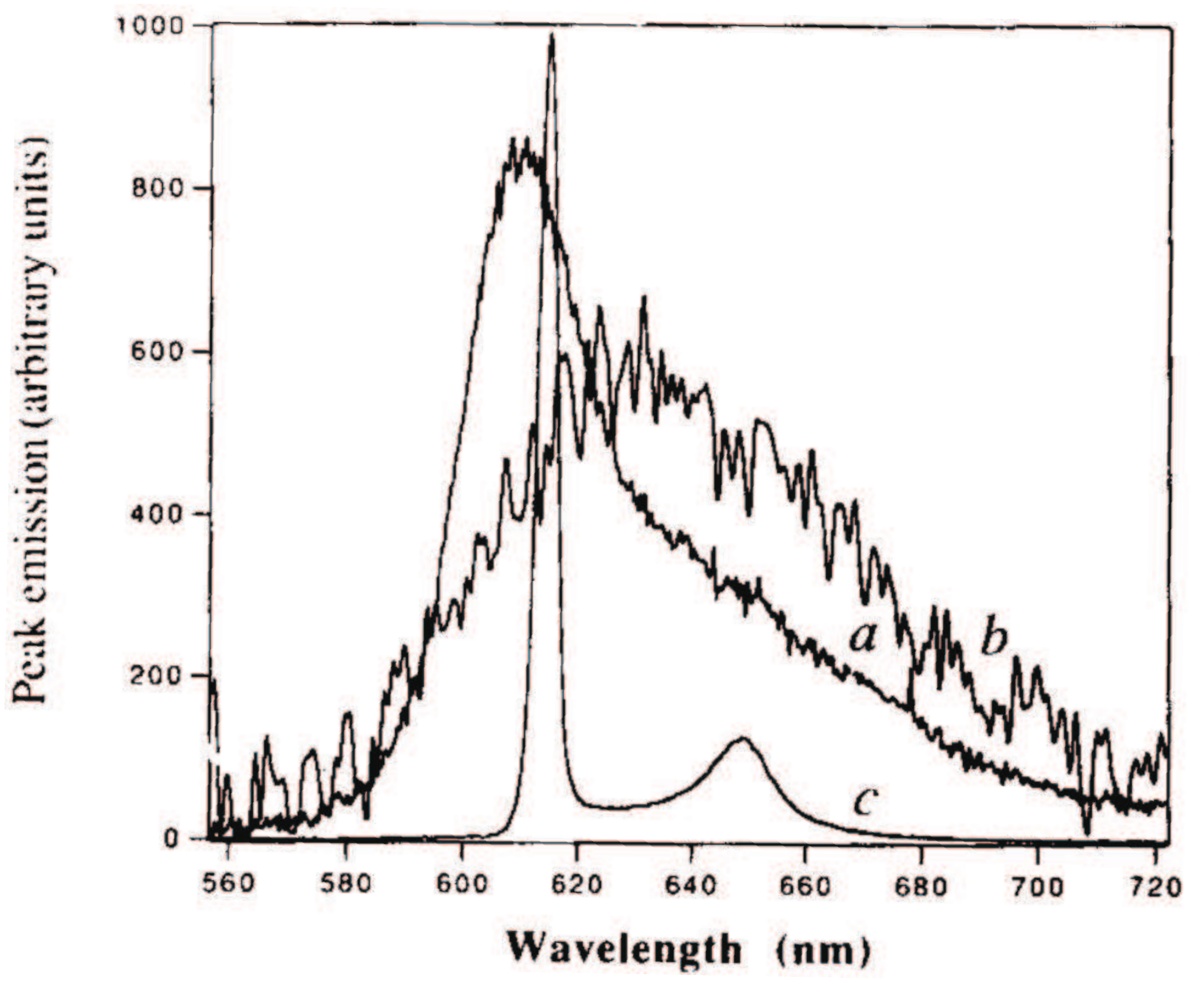}
\caption{(a) Emission spectrum of a $2.5\times 10^{-3}$ M solution of R640 perchlorate in methanol pumped by a 3-mJ (7 ns) pulses at 532 nm. (b) and (c) Emission spectra of the TiO$_2$ particles ($2.8 \times 10^10$ cm$^{-3}$ colloidal dye solution pumped by 2.2 $\mu$J and 3.3 mJ pulses, respectively. Emission: (b) scaled up 10 times, (c) scaled down 20 times. (Ref. \cite{104})}\label{Fig14}
\end{figure}

The lasing threshold is typically not suppressed substantially below the threshold for amplified stimulated emission in a neat dye solution. Light penetrates a depth into the sample equal to the absorption length of the pump radiation $L_a=\sqrt{D\tau_a}=\sqrt{\ell\ell_a/3}$ which is the same as the exponential falloff with thickness of transmission in diffusing systems with loss \cite{106}. Here $1/\tau_a$ is the absorption rate, $\ell$ the transport mean free path, $\ell_a$ = $v\tau_a$ the length of the trajectory in which the intensity falls to $1/e$ due to absorption, and $v$ is the transport velocity \cite{107}.
The excitation region illustrated in Fig. 15 is near the boundary so that the typical length of the paths of emitted photons would be comparable to those of the pump photons of $\sim \ell_a$ \cite{108}. But this is the length over which stimulated emission occurs in a neat solution. So the lasing threshold may not be lowered below the value at which appreciable amplified spontaneous emission occurs in a neat dye solution. The lasing threshold could be lowered by increasing the residence time inside the medium in samples with a shorter mean free path at the emission frequency than at the pump frequency or by internal reflection at the boundary. Above threshold, the optical transition pumped may be saturated so that absorption is suppressed and the wave can penetrate deeper into the sample.
\begin{figure}[htc]
\centering
\includegraphics[width=2in]{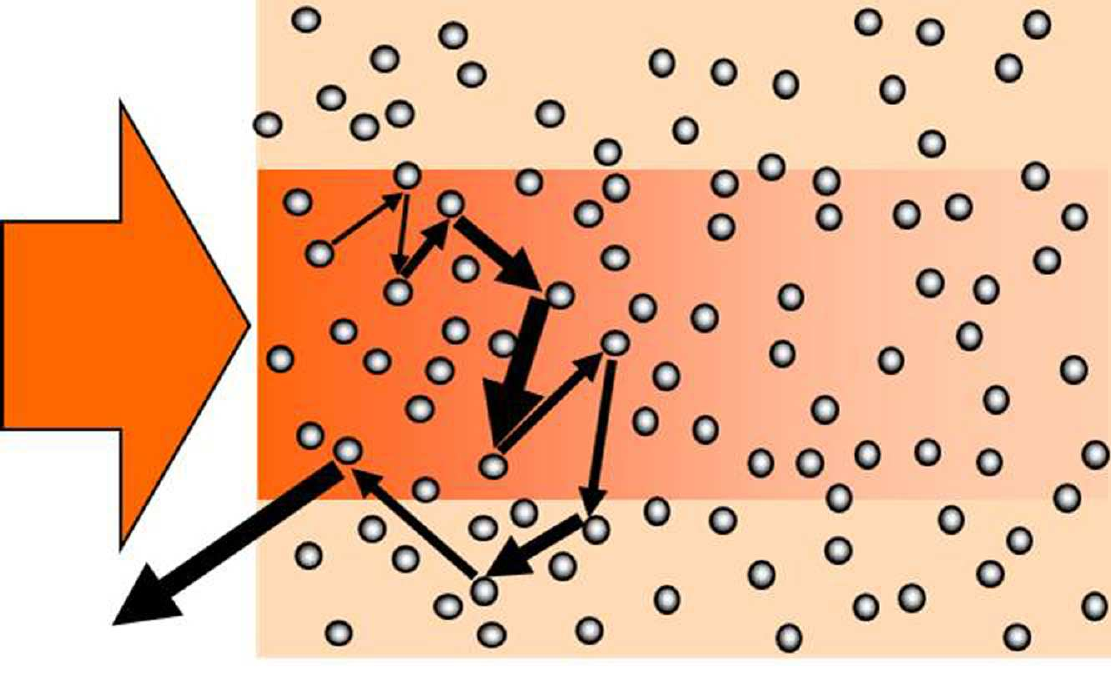}
\caption{Possible path for a photon emitted and amplified within a dye medium containing random scatterers. The lighter region indicates the volume pumped by the laser. (Ref. \cite{108})}\label{Fig15}
\end{figure}

The lasing threshold can be dramatically suppressed, however, for localized waves. When $\delta<1$, the intensity within the sample may grow exponentially when on resonance with a localized state far from the boundary. The excited region may then be in the middle of the sample and so emission will be into modes that overlap the excited mode and are similarly peaked in the middle of the sample and so are long lived. This was demonstrated in low-threshold lasing excited by a beam incident normally upon stacks of glass cover slips with thickness of approximately 100 $\mu m$ and intervening air layers of random thickness and with Rhodamine 6G dye solution between some of the slides \cite{45}. A plane wave incident upon parallel layers of random thickness is a one-dimensional medium and will be localized in the medium. In the present circumstance, the layers are not perfectly parallel and so light is scattered off the normal. This leads to a delocalization transition with a crossover at a thickness at which the transverse spread of an incident ray is equal to the size of the speckle spots formed \cite{64}. Beyond this thickness, the sample becomes three dimensional with regard to propagation of the initially normally incident beam. The spread of the beam is abetted by the thick layers used and could be reduced dramatically if layers with thickness $\sim \lambda/4$ were used.

Emission spectra excited by a pulsed Nd:YAG laser at 532 nm in a stack of cover slides with intervening dye solution recorded with a 0.07-nm-resolution grating spectrometer are shown in Fig. 16. The broad emission spectrum of the neat dye solution in Fig. 16(a) is compared with the emission spectrum from the random stack with interspersed dye layers slightly below and above the lasing threshold. Below threshold, the spectrum shows resolution-limited peaks of the electromagnetic modes of the system. Above threshold, a collimated emitted beam perpendicular to the sample layers was observed. The lasing spectrum with baseline shifted up for clarity is shown in Fig. 16(b). An abrupt change in the output power with increasing pump energy occurs at the lasing transition seen in the inset in Fig. 16(b). Just above threshold, lasing occurs in a single narrow line (Fig. 16(b)), while at higher energies, multimode lasing was observed (Fig. 16(c)) with wavelength and intensity that vary randomly with the position of the pump beam on the sample surface. The lasing threshold was low enough so that lasing could be observed with a chopped continuous wave Argon-ion laser beam at 3W at 514.5 nm.
\begin{figure}[htc]
\centering
\includegraphics[width=3in]{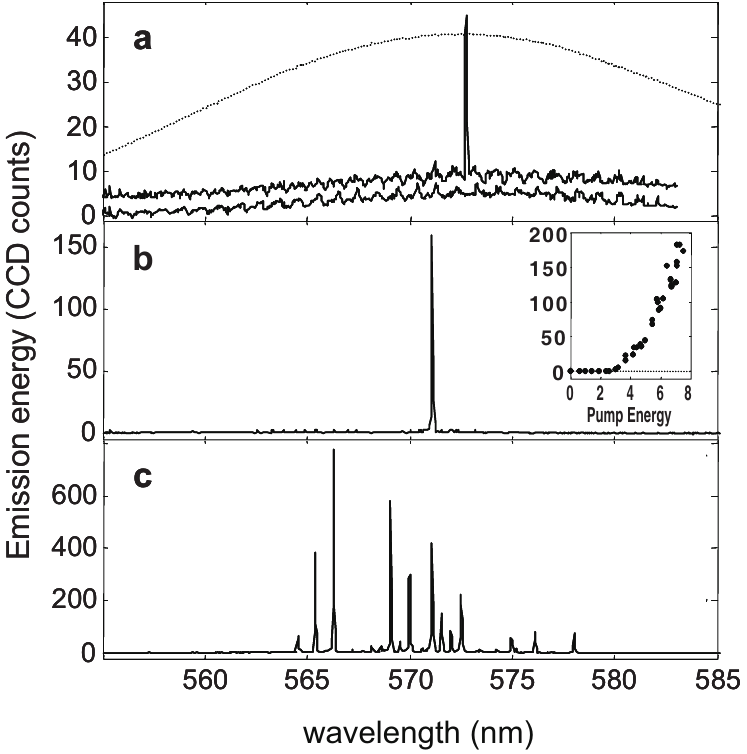}
\caption{(a) Spectra of spontaneous emission in neat solution and of spontaneous emission and near threshold lasing from Rhodamine 6G placed between layers of a glass slide stack at different laser pump energies.(b) Lasing in a single line above threshold and (c) in multiple lines well above threshold. The inset in (b) shows the sharp onset of lasing above threshold for excitation at a particular portion of the glass stack. (Ref. \cite{45})}\label{Fig16}
\end{figure} 

The role of resonance with localized modes at both the pump and emission wavelengths is seen in the strong correlation of pump transmission and output laser power. Such strong correlation is opposite to what would be expected for a nonresonant random laser in which peak emission would correspond to maximal absorption and so with reduced transmission. 

Low-threshold lasing via emission into long-lived modes excited by a pump laser which penetrates deeply into a sample can be realized in periodic and nearly periodic structures. For 1D samples or for layered structures in which the dielectric function is modulated only along a single direction, stop band are seen in the transmission spectrum perpendicular to layers. This is the case even when the layers are anisotropic with an orientation that varies with depth. The states at the edge of the band are long lived and can be excited via emission from excited states of dopants in the periodic structure or of the structure itself when pumped by an external beam falling within the frequency range of the pass band. A coherent beam perpendicular to the layers then emerges without special alignment.
 
In an infinite structure, the group velocity vanishes as the band edge is approached. This leads to the expectation of a lowered lasing threshold at the edge of a photonic band gap \cite{109}. But in periodic structures of finite thickness, states at the band edge are standing Bloch waves with a low number $n$ of anti-nodes in the medium rather than traveling waves \cite{110}. The intensity of the wave in each of these states is modulated by an envelope function $\sin^2n\pi x/L$, where $x$ is the depth into the sample of thickness $L$. Lasing properties are determined by the modes with increasingly narrow linewidths and intensity as the band edge is approached. The width of modes increases as $n^2$ away from the band gap.

Band edge lasing was demonstrated in dye-doped cholesteric liquid crystals (CLCs) \cite{44}. Lasing from dye-dope CLCs was observed earlier and attributed to lasing at defect sites in the liquid crystal \cite{111}. Roughly parallel rod-shaped molecules in CLCs with average local orientation of the long molecular axis in a direction called the director rotate with increasing depth into the sample. This periodic helical structure can be either right- or left-handed. The indices for light polarized parallel and perpendicular to the director are the extraordinary and ordinary refractive indices, $n_e$ and $n_o$, respectively. For sufficiently thick films, the reflectance of normally incident, circularly polarized light with the same sign of rotation as the CLC structure is nearly complete within a band centered at vacuum wavelength $\lambda_c=nP$ where $n = (n_e+n_o)/2$ and $P$ is the pitch of the helix equal to twice the structure period. The reflected light has the same sign of rotation as the incident beam. The bandwidth is $\Delta \lambda=\lambda_c\Delta n/n$, where $\Delta n= n_e-n_o$. For circularly polarized light of opposite circular polarization, the wave is freely transmitted. 
\begin{figure}[htc]
\centering
\includegraphics[width=2in]{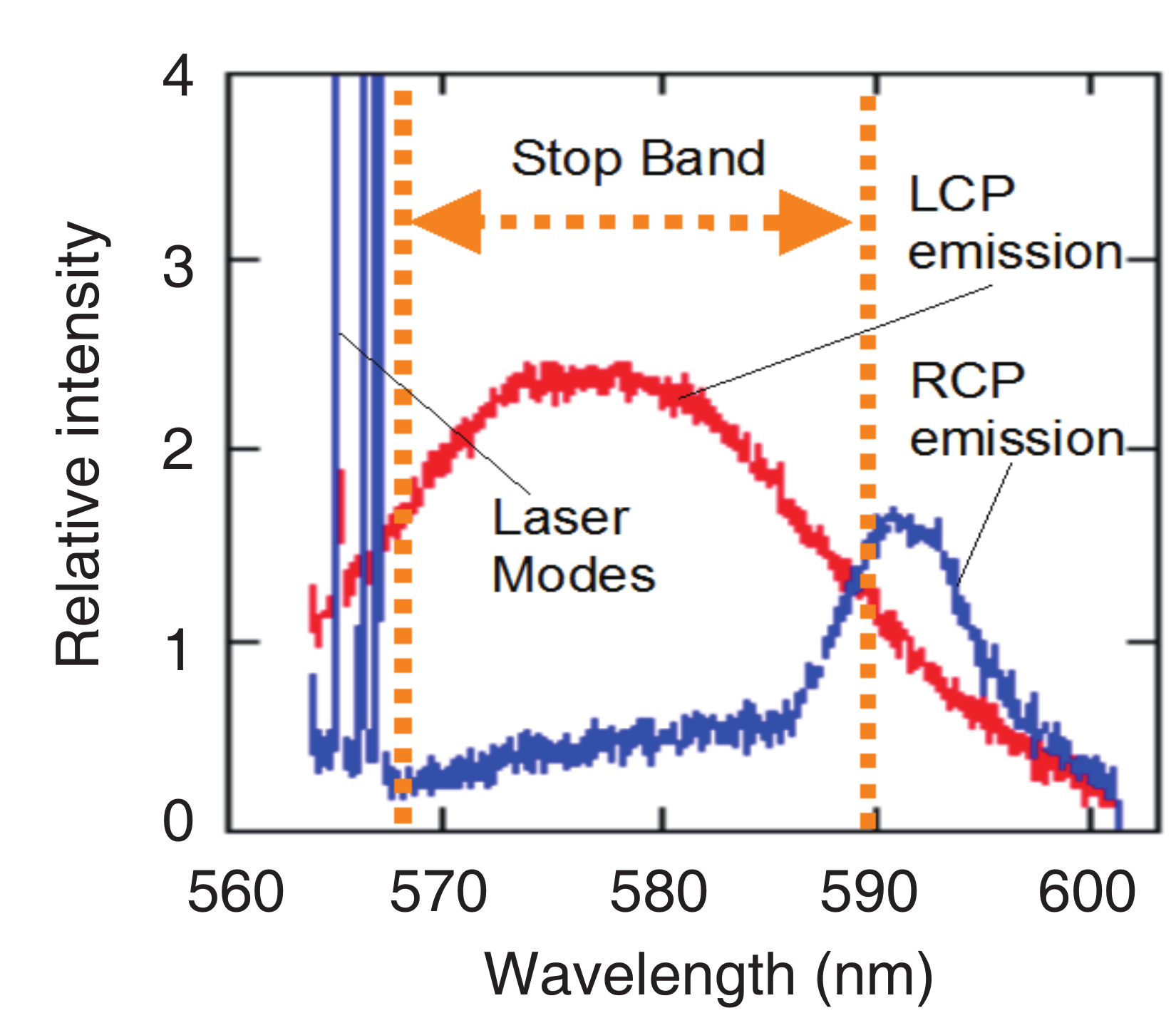}
\caption{Left and right circularly polarized emission spectra from a right handed dye-doped CLC sample as well as lasing emission at the short-wavelength edge of the reflection band. The height of the lasing lines is $\sim$50 units. (Ref. \cite{44})}\label{Fig17}
\end{figure} 
In measurements on dye-doped cholesteric liquid crystal (CLC) films, spontaneous emission is inhibited within the band and the density of states is enhanced at the band edge for light polarized with the same handedness as the chiral structure. Light of opposite chirality is unaffected by the periodic structure. This makes it possible to make a direct measurement of the density of photon states by comparing the emission spectra of oppositely polarized radiation. The observed suppression of the density of states within the band and the sharp rise at the band edge are shown in Fig. 17 and seen to be in good agreement with the calculated density of states in a 1D structure. The left circularly polarized (LCP) emission spectrum in this right handed structure is due to the spontaneous emission of the PM-597 dye. RCP emission is suppressed in the stop band and peaked at the band edges. The RCP light seen within the reflection band does not vanish because the emitted LCP light is converted to RCP light in Fresnel reflection from the surfaces of the glass sample holder. Multiple lasing lines are seen at the short-wavelength band edge. 

The lasing peaks in Fig. 17 do not correspond precisely to the modes of a perfectly periodic CLC structure. These modes are seen in transmission spectra in Fig. 18 in a dye doped CLC sample which was carefully prepared and allowed to equilibrate. A comparison between transmission measured with a tunable narrowband dye laser in a 37-$\mu m$ thick CLC sample with moderate absorption and simulations for a periodic system is shown below. The simulated spectrum is displaced vertically for visibility. In a nondissipative sample, the resonance transmission of all modes modes reaches unity. In Fig. 18, transmission through modes closest to the band edge is most suppressed by absorption since these modes are longest lived. Since the modes closest to the band edge are longest lived in nearly periodic systems, these states are most susceptible to being localized by disorder. Such localized states are often longer lived than the corresponding states of a periodic system and so disorder can help as well as hinder lasing.
\begin{figure}[htc]
\centering
\includegraphics[width=2.5in]{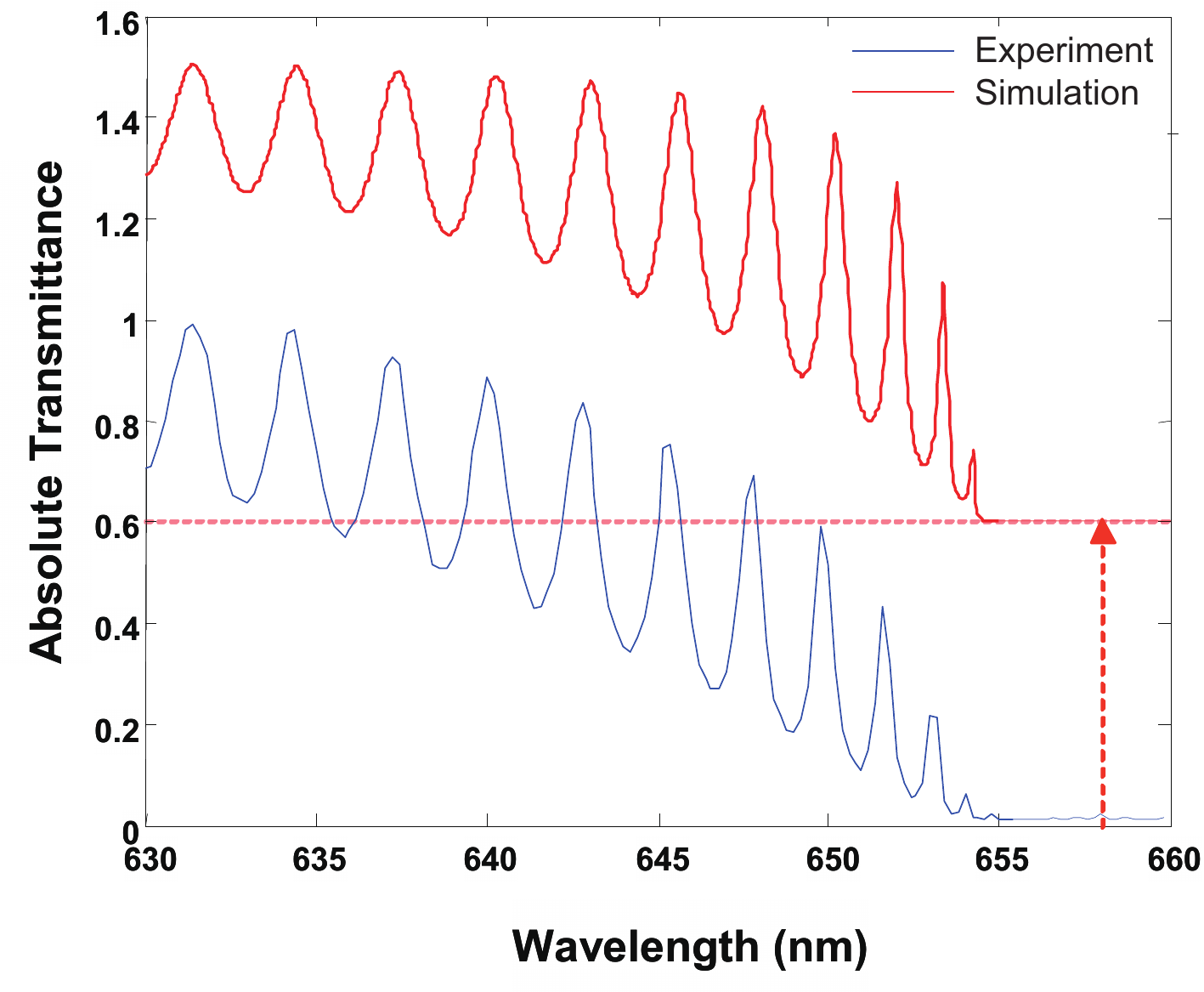}
\caption{Comparison of measurements of transmission spectra in dye-doped CLC taken by Valery Milner in the lower curve with simulations in the upper curve. The linewidth narrows as the index $n$ of the mode away from the band edge decreases. Differences between the frequencies of laser lines seen in Fig. 17 and frequencies of lines in high quality CLC samples in this figure are due to disorder in the sample of Fig. 17.}\label{Fig18}
\end{figure}

Simulations in random amplifying systems show that it is possible to maximize the lasing intensity at a particular frequency in the spectrum of a random laser by iteratively feeding back the intensity at a selected frequency to vary the intensity distribution of the pump beam \cite{112}. The modes of the sample are not substantially modified in the lasing transition, but the spectral properties of the modes excited by the pump beam are selected by the spatial profile of the pump beam. T\"ureci {\it et al.} have shown that modes of passive diffusive systems interact via the gain medium to create a uniform spacing in the laser spectrum \cite{113}. In contrast, isolated modes of localized lasers interact weakly and emit at a frequencies pegged to the modes of the passive systems \cite{113a}.

\section{Channels}
Transmission through a disordered medium is fully determined by the transmission matrix $t$ \cite{11,46,52,53}. The optical transmission matrix was measured by Popoff {\it et al.} \cite{41} with use of a spatial light modulator (SLM) and an interference technique to find the amplitude and phase of the optical field. Measuring the transmission matrix allows one to focus the transmitted light at a desired channel at the output surface by phase conjugating the transmission matrix \cite{41,48}. In this way, the transmitted field from different input channels arrives in phase at the focal spot and interferes constructively. The presence of a random medium can increase the number of independent channels that illuminate a point so that the focused intensity and resolution are enhanced \cite{115,116,117}. Because of the enormous number of channels in optical experiments, only a small portion of the transmission matrix is typically measured. The distribution of the singular values of the transmission matrix may then follow the quarter circle law which is characteristic of uncorrelated Gaussian fluctuations of the elements of the transmission matrix \cite{118,119}.

Measurements of microwave radiation propagation through random media confined in a waveguide allow us to measure the field on a grid of points for the source and detector \cite{42}. The closest spacing between points is approximately the distance at which the field correlation function vanishes so that the fields at different points on the gird are only weakly correlated. The number of independent channels $N$ supported in the empty waveguide is $\sim 66$ in the frequency range of 14.7-14.94 GHz in which the wave is diffusive and $\sim 30$ from 10-10.24 GHz in which the wave is localized within the sample. To construct the transmission matrix, $N$/2 points are selected from each of two orthogonal polarizations. A representation of intensity patterns in typical transmission matrices for both diffusive and localized waves at a given frequency is presented in Fig. 19. Each column presents the variation of intensity across the output surface at points $b$ for a source at points $a$ with two orthogonal polarizations. The intensity in each column shown in Fig. 19 is normalized by its maximum value. For localized waves, intensity patterns in each column are similar indicating that transmission is dominated by a single channel. In contrast, no clear pattern is seen for diffusive waves since many channels contribute to the intensity at each point.
\begin{figure}[htc]
\centering
\includegraphics[width=2.5in]{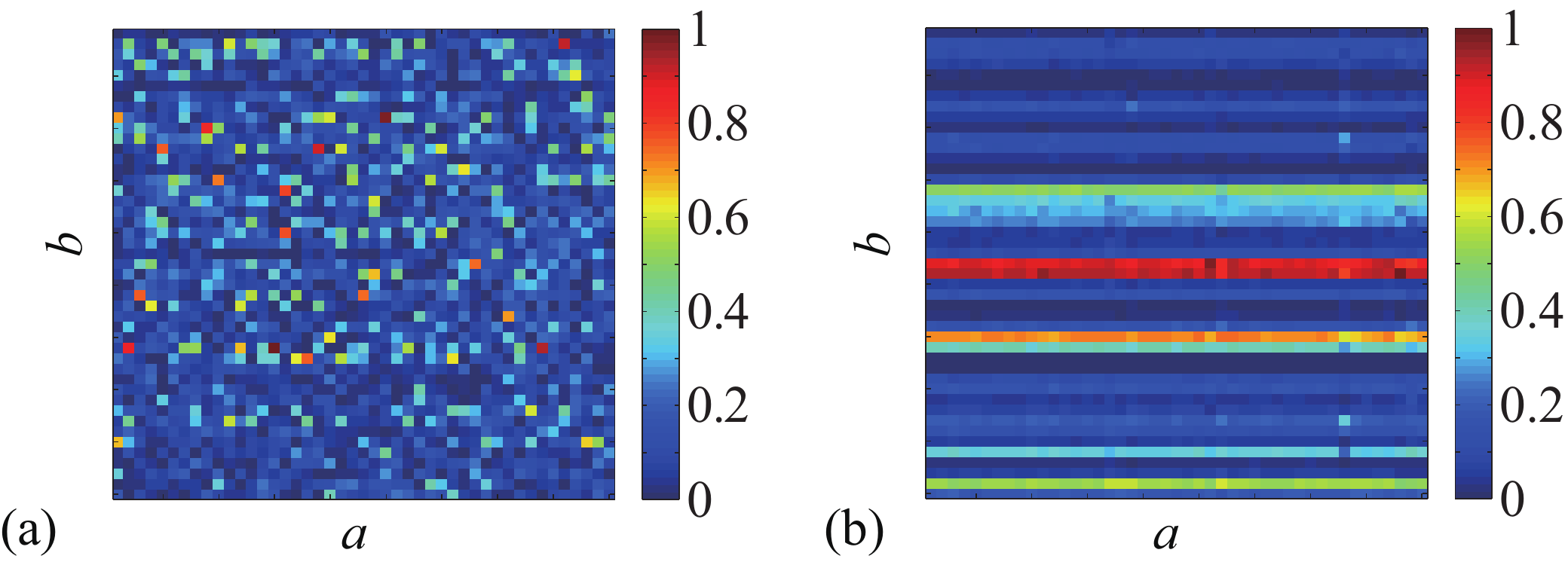}
\caption{Intensity normalized to the peak value in each speckle pattern generated by sources at positions $a$ are represented in the columns with index of detector position and polarization $b$ for (a) diffusive and (b) localized waves. (Ref. \cite{47})}\label{Fig19}
\end{figure} 

In Fig. 20, we show a spectrum of the optical transmittance and the underlying transmission eigenvalues from a single random realization for both localized and diffusive waves. This confirms that the highest transmission channel dominates the transmittance for localized waves while several channels contribute to transmission for diffusive waves. Thus for localized waves, the incident wave from different channels couples to the same eigenchannel and excites the same pattern in transmission as seen in Fig. 19(b). In contrast, the transmission patterns for incident waves for different incident channels are the sums of many orthogonal eigenchannels so that the transmitted patterns are weakly correlated.
\begin{figure}[htc]
\centering
\includegraphics[width=3.5in]{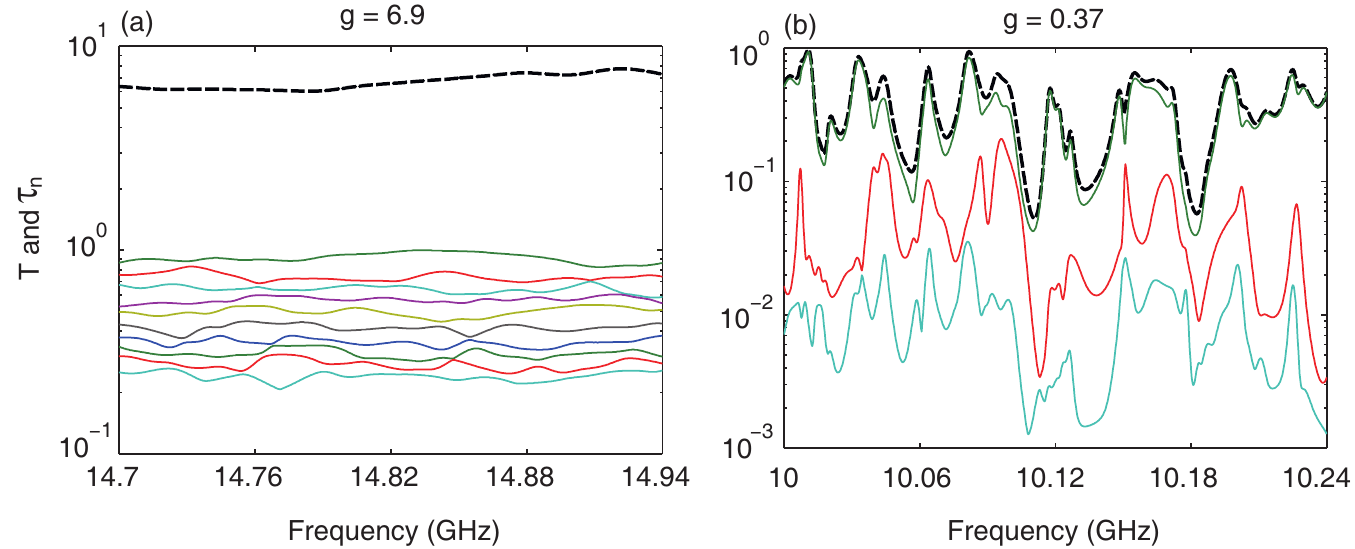}
\caption{Spectra of the transmittance $T$ and transmission eigenvalues $\tau_n$ for (a) diffusive sample of $L=23$ cm with {\textsl g}=6.9 and (b) localized sample of $L=40$ cm with {\textsl g}=0.37. The black dashed line gives $T$ and the solid lines are spectra of $\tau_n$. (Ref. \cite{42})}\label{Fig1}
\end{figure} 

Dorokhov \cite{10,11} showed that, the spacing between the inverse of the localization length for adjacent eigenchannels is equal to the inverse of the localization length of the sample, $\frac{1}{\xi_{n+1}}-\frac{1}{\xi_n}=\frac{1}{\xi}$. For localized waves, this is equivalent to $\langle \ln \tau_n\rangle -\langle \ln \tau_{n+1}\rangle=L/\xi=1/{\textsl g}_0$, where $\textsl{g}_0$ is the bare conductance that one would obtain in the absence of wave interference and the transport can be described in terms of diffusion of particles. In Fig. 21, we show that $\langle \ln \tau_n \rangle$ falls linearly with respect to the channel index $n$ for both diffusive and localized waves. We denote the constant spacing between adjacent values of $\langle \ln \tau_n \rangle$ as $1/{\textsl g}^{\prime\prime}$, $\langle \ln \tau_n\rangle -\langle \ln \tau_{n+1}\rangle=1/{\textsl g}^{\prime\prime}$. This supports the conjecture that ${\textsl{g}^{\prime\prime}}$ is the bare conductance.
\begin{figure}[htc]
\centering
\includegraphics[width=2in]{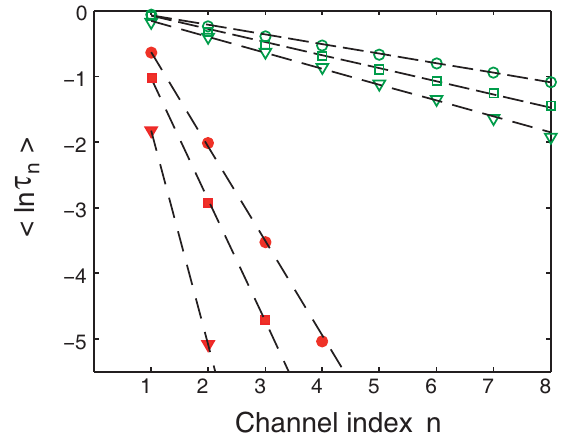}
\caption{Variation of $\langle \ln \tau_n \rangle$ with channel index $n$ for sample lengths $L=23$ (circle), 40 (square) and 61 (triangle) cm for both diffusive (green open symbols) and localized (red solid symbols) waves fitted, respectively, with black dashed lines. (Ref. \cite{42})}\label{Fig21}
\end{figure}  

We expect that the bare conductance should be influenced by the wave interaction at the sample interface \cite{120,121,122}. The wave interaction at the sample boundary can be described by a diffusion model \cite{122} in which the incident wave is replaced by an isotropic source at a distance $z_p$ from the interface in which the wave direction is randomized and with a length $z_0$, which is the length beyond the sample boundary at which the intensity inside the sample extrapolates to zero. $z_0$ was found by fitting the time of flight distribution of wave through random media \cite{80, 123}. Once the surface effect is taken into account, the bare conductance is given as: ${\textsl g}=\eta\xi/L_{eff}$, where $\eta$ is a constant of order of unity and $L_{eff} = L+2z_0$ is the effective sample length. The constant value of ${\textsl g}^{\prime\prime}L_{eff}$ seen in Fig. 22 is consistent with ${\textsl g}^{\prime\prime}$ being the bare conductance and gives the localization length for the samples at two frequency ranges. The absolute values of the transmittance $T$ and of the underlying transmission eigenvalues $\tau_n$ are obtained by equating $\langle T\rangle=C{\textsl g}^{\prime\prime}$ for the most diffusive sample of length $L = 23$ cm, at which the renormalization of dimensionless conductance due to wave localization is negligible. The normalization factor $C$ is used to determine the values of {\textsl g} for other samples.
\begin{figure}[htc]
\centering
\includegraphics[width=2in]{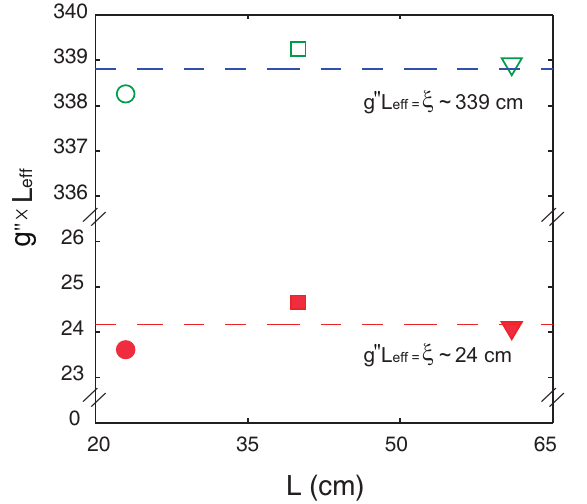}
\caption{The constant products of ${\textsl g}^{\prime\prime}L_{eff}$ for three different lengths for both diffusive and localized samples give the localization length $\xi$ in the two frequency ranges. (Ref. \cite{42})}\label{Fig22}
\end{figure} 

The probability density of $\ln \tau_n$ of the first few eigenchannels and their contribution to the overall density $\ln \tau$ is shown in Fig. 23 for the most diffusive sample with {\textsl g}=6.9. 
\begin{figure}[htc]
\centering
\includegraphics[width=2.5in]{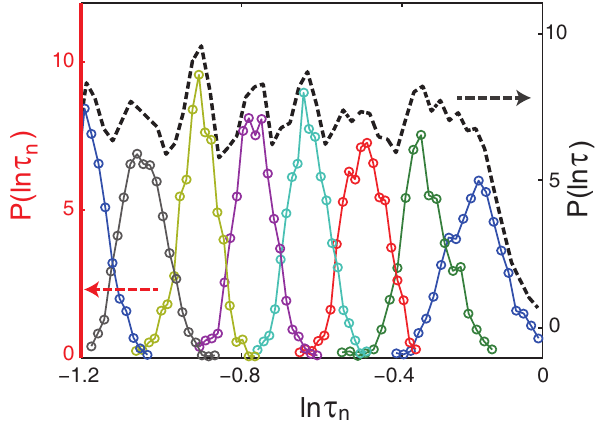}
\caption{Probability density of $\ln \tau_n$ (lower curves) and the density of $\ln \tau$ (top dashed curve), $P(\ln \tau)=\sum_{n}{P(\ln \tau_n)}$ for the diffusive sample with {\textsl g}=6.9. (Ref. \cite{42})}\label{Fig23}
\end{figure} 
Aside from the fall of the probability distribution $P(\ln \tau)$ near $\ln \tau \sim 0$, which reflects the restriction $\tau_1\leq 1$, $P(\ln \tau)$ is nearly constant with ripples spaced by $1/{\textsl g}^{\prime\prime}$. The nearly uniform density $P(\ln \tau)$ of corresponds to a probability density $P(\tau)=P(\ln \tau)\frac{d\ln \tau}{d\tau}={\textsl g}/\tau$. This distribution has a single peak at low values of $\tau$, in contrast to the predicted bimodal distribution, which has a second peak nearly unity \cite{11,46,52,124}. This may reflect the fundamental difference of measuring the transmission matrix based on scattering between independent discrete points instead of waveguide modes. In theoretical calculation in which scattering between waveguide modes is treated, all the transmitted energy can be captured. However, only a fraction of energy transmitted through the disordered medium is captured when the TM is measured on a grid of points. As a result, full information is not available and the measured distribution of transmission eigenvalues does not accurately represent the actual distribution in the medium. In particular the bimodal distribution of transmission eigenvalues is not observed. This has been suggested in recent simulation of a scalar wave propagation in Q1D samples based on recursive Green’s function method. Goetschy and Stone \cite{126} have recently calculated the impact of degree of control of the transmission channels on the density of transmission eigenvalues. The matrix $t$ is mapped to $t^\prime=P_2 t P_1$, where $P_1$, $P_2$ are $N\times M_1$ and $M_2\times N$ matrices which eliminate $N-M_1$ columns and $N-M_2$ rows, respectively, of the original random matrix $t$. Therefore, only $M_1$($M_2$) channels  are under control on the input (output) surface, respectively, and the degree of control on the input and output surfaces is measured by $M_1/N$ ($M_2/N$). As a result, the density of transmission eigenvalues for diffusive samples changes from a bimodal distribution to a distribution characteristic of uncorrelated Gaussian random matrices, when the degree of control is reduced \cite{41,42,127}. Nevertheless, key aspects of the statistics of wave propagation and the limits of control of the transmitted wave can be explored using measurements of the transmission matrix.

Measuring the transmission matrix allows us to explore the statistics of transmittance, the most spatially averaged mesoscopic quantity. The importance of sample-to-sample fluctuation of conductance in disordered conductors was first recognized in conduction mediated by localized states, but fluctuations in transmission were first observed in the constant variance of fluctuations of conductance in diffusive samples known as universal conductance fluctuations \cite{9,56,70a,70,128}. For diffusive waves, for which a number of transmission eigenchannels contribute substantially to the transmittance, the probability distribution of $T$ is Gaussian with variance independent of the mean value of $T$ and of sample dimensions. In the localization limit, $L/\xi\gg 1$, in which transmittance $T$ is dominated by the largest transmission eigenvalues, $T\sim \tau_1$, the single parameter scaling (SPS) theory of localization predicts that the probability distribution of the logarithm of transmittance in 1D samples is a Gaussian function with a variance equal to the average of its magnitude, var($\ln T$)=$-\langle \ln T\rangle$. Therefore, the scaling of average of conductance and the entire distribution of conductance is determined by the single parameter $|\langle \ln T\rangle|/L=1/\xi$. In recent work, the ratio $\mathscr{R}\equiv -{\textrm var}(\ln T)/\langle \ln T\rangle$, is found to approach unity in Q1D samples showing that propagation in Q1D in this limit is one-dimensional \cite{69}. 

In the Q1D geometry, there is no phase transition between localization and diffusion as {\it L} increases for samples with equivalent local disorder. Instead, there exists a crossover from the diffusive to localized regime. For samples just beyond the localization threshold, in which only a few transmission eigenchannels contribute appreciably to the transmittance, numerical simulation \cite{129,130,131,132,133} and random matrix theory calculation \cite{134} by Muttalib and W\"olfle found a one-sided log-normal distribution for the transmittance. The source of this unusual probability distribution of conductance can be understood with the aid of the charge model proposed by Stone, Mello, Muttalib, and Pichard \cite{46}. The charge model was first introduced by Dyson \cite{135} to visualize the repulsion between eigenvalues of the large random Hamiltonian. In this model, transmission eigenvalues $\tau_n$ are associated with positions of parallel line charges at $x_n$ and their images at $-x_n$ embedded in a compensating continuous charge distribution. The transmission eigenvalues are related to the $x_n$ via the relation, $\tau_n=1/\cosh^2x_n$. The repulsion between two parallel lines of charges of the same sign with potential $\ln|x_i-x_j|$ mimics the interaction between eigenvalues of the random matrix. The oppositely charged jellium background provides an overall attractive potential that holds the structure together. The repulsion between charges for diffusive waves is the origin of universal conductance fluctuation. For localized waves, the charges are separated by a distance greater than the screening length due to the background charges so that the ``Coulomb" interaction is screened. The repulsion between the first charge at $x_1$ associated with the highest transmission eigenvalues $\tau_1$ and its image placed at $-x_1$ provides ceiling of unity.

We have recently reported microwave measurements of the probability distribution of the``optical" transmittance {\it T} in the crossover from diffusive to localized waves \cite{69}. A Gaussian distribution is found for diffusive waves and a nearly log-normal distribution for deeply localized waves. Just beyond the localization threshold, a one-sided log-normal distribution is observed for an ensemble with {\textsl g}=0.37. In this ensemble, an exponential decay of $P(T)$ is found for high values of transmittance as was found in simulations and calculations \cite{136}. The rapid falloff of $P(T)$ for $T>1$ is due to the requirement that two eigenvalues need to be high in this case. This requires that two charges as well as their images be close to the origin. The probability for high values of $T$ is therefore greatly suppressed due to the repulsion between these charges. 

Measurements of the transmission matrix provide the opportunity to investigate the statistics in single disordered samples as opposed to the statistics of ensembles of random sample. Such statistics are essential in applications such as imaging and focusing through a random medium. In the Q1D geometry, in which the wave is completely mixed within the sample, the statistics of the intensity relative to the average over the transmitted speckle pattern, $T_{ba}/(\sum_{b=1}^N{T_{ba}/N})=NT_{ba}/T_a$, is independent of source or detector positions \cite{137,138}. Because of the Gaussian distribution of the field in any single speckle pattern, the probability distribution of relative intensity is $P(NT_{ba}/T_a)=\exp(-NT_{ba}/T_a)$. Since the statistics of relative intensity are universal, the statistics of the transmission in a sample with transmittance $T$ would be completely specified by the statistics of total transmission $T_a$ relative to its average $(T/N)$ within the sample. 

We find in random matrix calculations that the variance of normalized total transmission within a single instance of a large transmission matrix is equal to the inverse eigenchannel participation number \cite{47},
\begin{equation}
{\textnormal var}(NT_a/T)=M^{-1}.
\end{equation}
These results can be compared to measurements in samples of small $N$ by grouping together measurements in collections of samples with similar values of $M$. We show in Fig. 24 that the average of var($NT_a/T$) in subsets of samples with given $M^{-1}$ is in excellent agreement with Eq. 2. var[var($NT_a/T$)/$M^{-1}$] is seen in the insert of Fig. 24 to be proportional to $1/N$ indicating that fluctuations in the variance over different subsets are Gaussian with a variance that vanishes as $N$ increases.
\begin{figure}[htc]
\centering
\includegraphics[width=2.5in]{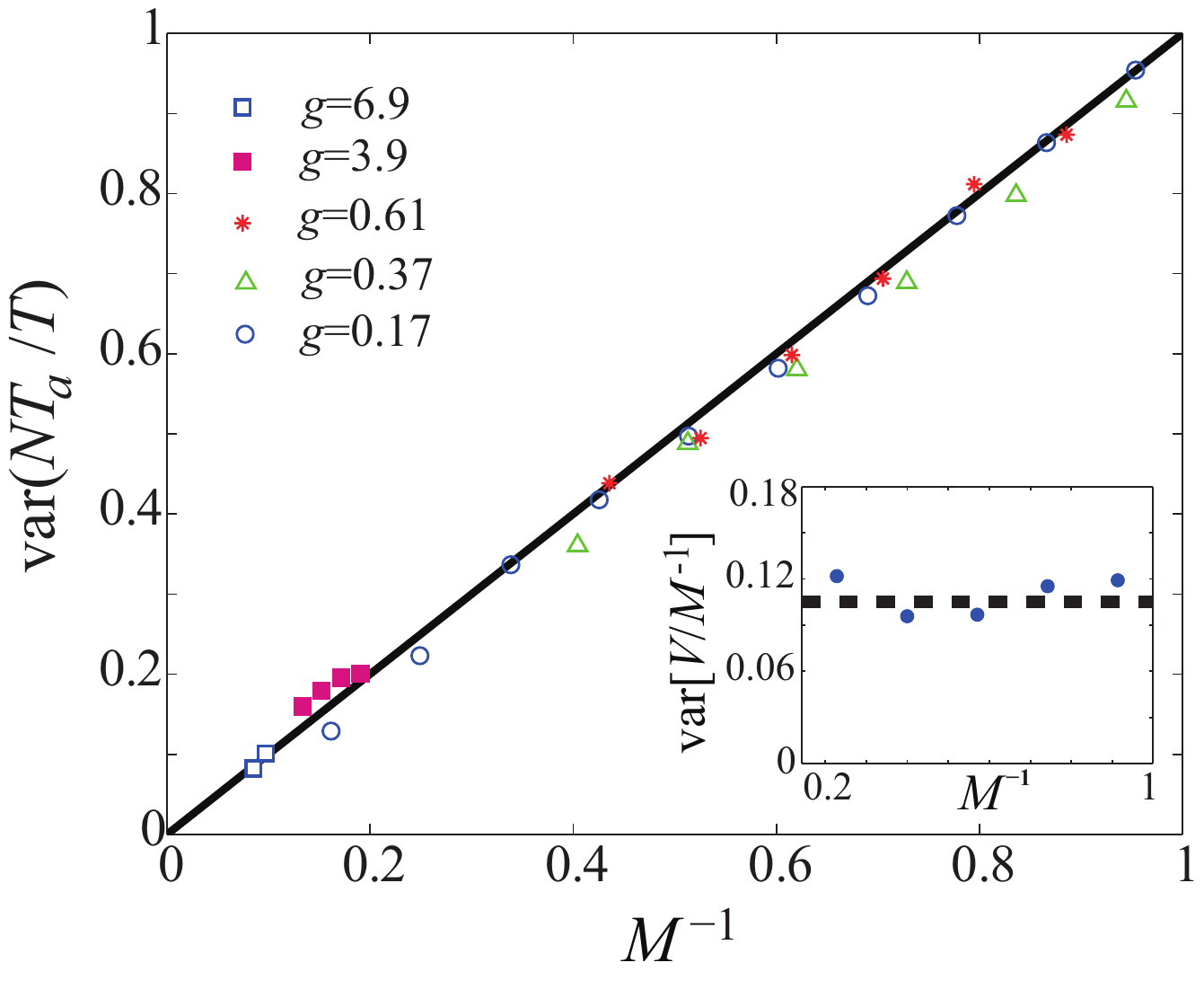}
\caption{Plot of the var$(NT_a/T)$ computed within transmission matrices over a subset of transmission matrices with specified value of $M^{-1}$ drawn from random ensembles with different values of {\textsl g}. The straight line is a plot of var($NT_a/T$) = $M^{-1}$. In the inset, the variance of $V/M^{-1}$ is plotted vs. $M^{-1}$, where $V = {\textnormal var}(NT_a/T)$. (Ref. \cite{47})}\label{Fig24}
\end{figure} 

The central role played by $M$ can be appreciated from the plots shown in Fig. 25 of the statistics for subsets of samples with identical values of $M$ but drawn from ensembles with different values of {\textsl g}. The distributions $P(NT_a/T)$ obtained for samples with $M^{-1}$ in the range $0.17\pm0.01$ selected from ensembles with {\textsl g=3.9} and 0.17 are seen to coincide in Fig. 25(a) and thus to depend only on $M^{-1}$. The curve in Fig. 25(a) is obtained from an expression for $P(s_a)$ for diffusive waves given in Refs. \cite{30,31}, in terms of a single parameter {\textsl g}=2/3var($s_a$) but with the substitution of $2/3M^{-1}$ for {\textsl g}. The dependence of $P(NT_a/T)$ on $M^{-1}$ alone and its independence of $T$ is also demonstrated in Fig. 25(b) for $M^{-1}$ over the range $0.995\pm0.005$ from measurements in samples of different length with {\textsl g}=0.37 and 0.17. Since a single channel dominates transmission in the limit, $M^{-1}\rightarrow 1$, we have $NT_a/T=|v_{1a}|$, where $v_{1a}$ is the element of the unitary matrix $V$ which couples the incident channel $a$ to the highest transmission channel. The Gaussian distribution of the elements of $V$ leads to a negative exponential distribution for the square amplitude of these elements and similarly to $P(NT_{a}/T)=\exp(-NT_a/T)$, which is the curve plotted in Fig. 25. In Fig. 25, we plot the relative intensity distributions $P(N^2T_{ba}/T)$ corresponding to the same collection of samples as in Fig. 25, respectively. The curves plotted are the intensity distributions obtained by mixing the distributions for $P(NT_a/T)$ shown in Fig. 25 with the universal negative exponential function for the intensity of a single component of polarization.
\begin{figure}[htc]
\centering
\includegraphics[width=3in]{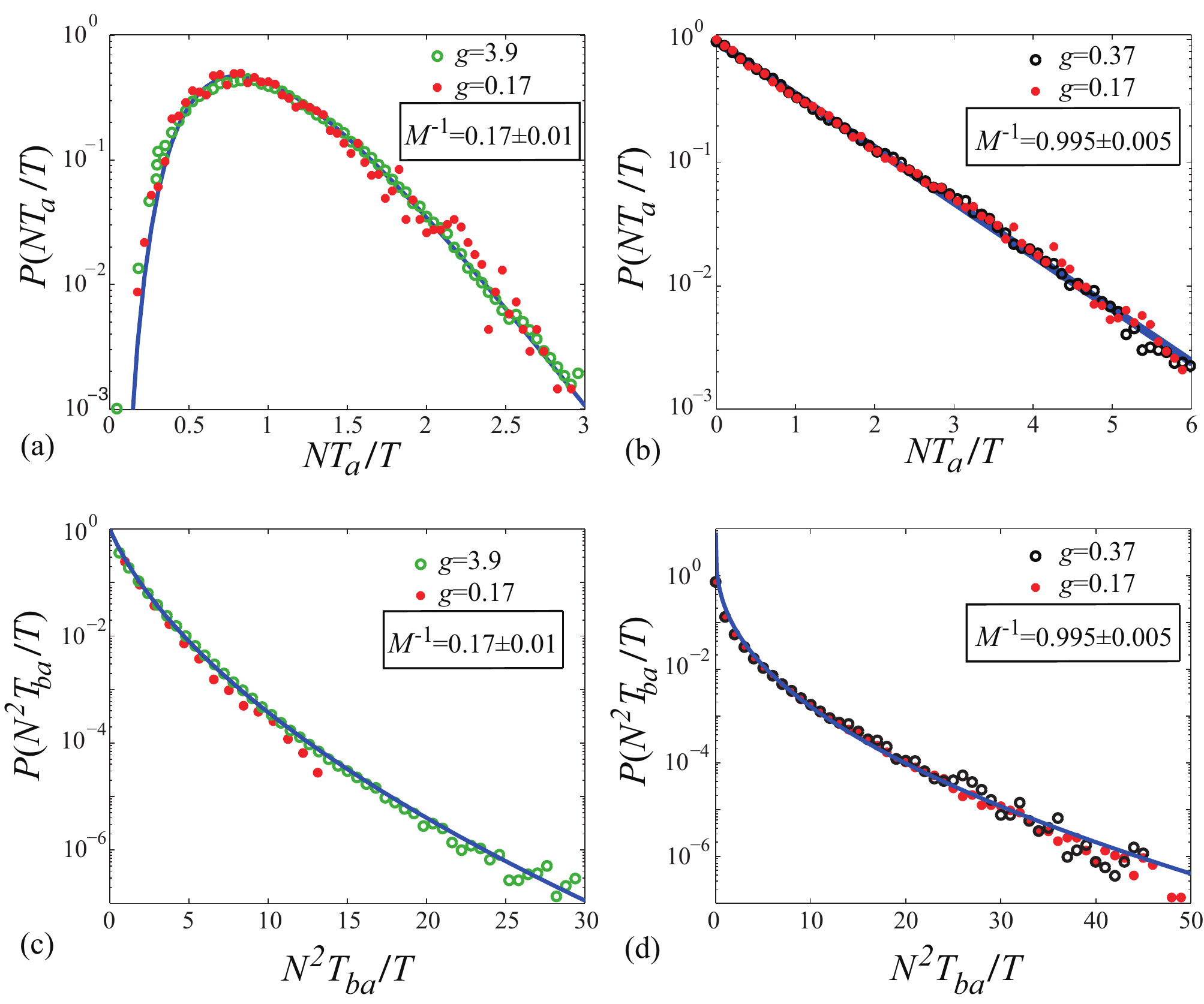}
\caption{(a) $P(NT_a/T)$ for subsets of transmission matrices with $M^{-1}$ = $0.17 \pm 0.01$ drawn from ensembles of samples with $L = 61$ cm in two frequency ranges in which the wave is diffusive (green circles) and localized (red filled circles). The curve is the theoretical probability distribution of $P(s_a)$ in which var$(s_a)$ is replaced by $M^{-1}$ in the expression for $P(NT_a/T)$ in Ref. \cite{30,31}. (b) $P(NT_a/T)$ for $M^{-1}$ in the range $0.995 \pm 0.005$ computed for localized waves in samples of two lengths: $L = 40$ cm (black circles) and $L = 61$ cm (red filled circles). The straight line represents the exponential distribution, $\exp(-NT_a/T)$. (c,d) The intensity distributions $P(N^2T_{ba}/T)$ are plotted under the corresponding distributions of total transmission in (a) and (b). (Ref. \cite{47})}\label{Fig25}
\end{figure} 

In addition, we find in microwave measurements in Q1D samples that the SPS ratio $\mathscr{R}$ is equal to average of $M$ weighted by {\it T}, ${\langle M T \rangle}/\langle T \rangle$, which approaches unity for $L/\xi \gg 1$ \cite{69}. The statistics of relative transmission within a single transmission matrix depends only upon the single parameter $M$ while the transmittance $T$ serves as an overall normalization factor. Therefore, the statistics of intensity and total transmission over random ensemble is given by the joint probability distribution of $T$ and $M$.

\section{Focusing}
Focusing waves through random media was first demonstrated in acoustics by means of time reversal \cite{139}. The amplitude and phase of the transmitted signal in time for an incident pulse from a source is picked up by arrays of transducers. The recorded signal is then played back in time and a pulse emerges at the location of the source. Recently, Vellekoop and Mosk \cite{140} focused monochromatic light through opaque media by shaping the incident wavefront. Employing a genetic algorithm with a feedback from the intensity at the target point to adjust the phase of the incident wavefront, the intensity at the focus was enhanced by three orders of magnitude. The wavefront shaping method has been extended to focus optical pulses through random media at a spatial target at selected time delay \cite{141,142}.

In order to focus a wave at a target channel $\beta$ once the field transmission matrix has been measured, one simply conjugates the phase of the incident field relative to the transmitted field at $\beta$, yielding $t_{\beta a}^*/\sqrt{T_\beta}$ for the normalized incident field. Here, the incident field is normalized by $T_\beta=\sum_{a=1}^N{|t_{\beta a}|^2}$ so that the incident power is set to be unity. In this way, the field from different incident channels $a$ arrives at the target in phase and interfere constructively. Random matrix calculations confirmed by microwave measurements show that the contrast between the average intensity at the focal spot $\langle I_\beta\rangle$ and the background intensity $\langle I_{b\neq \beta}\rangle$, $\mu=\langle I_\beta\rangle/\langle I_{b\neq \beta}\rangle$, depends upon the eigenchannel participation number $M$ and size of the measured transmission matrix $N$ \cite{47},
\begin{equation}
\mu=\frac{1}{1/M-1/N}.
\end{equation}
\begin{figure}[htc]
\centering
\includegraphics[width=2.5in]{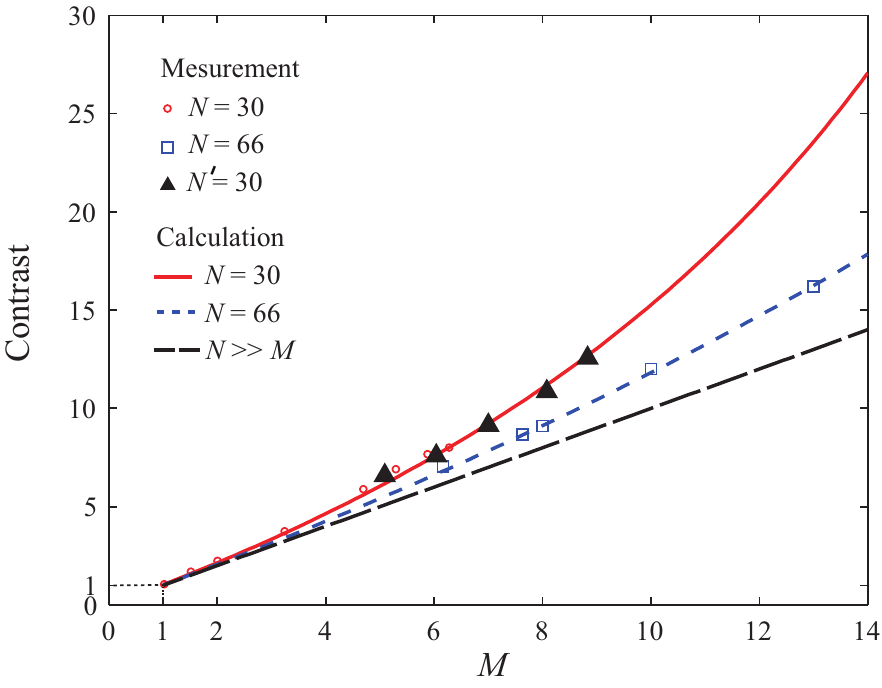}
\caption{Contrast in maximal focusing vs. eigenchannel participation number $M$. The open circles and squares represent measurements from transmission matrices $N = 30$ and 66 channels, respectively. The filled triangles give results for $N^\prime \times N^\prime$ matrices with $N^\prime = 30$ for points selected from a larger matrix with size $N = 66$. Phase conjugation is applied within the
reduced matrix to achieve maximal focusing. Equation (3) is represented by the solid red and dashed blue curves for $N = 30$ and 66, respectively. In the limit of $N\gg M$, the contrast given by Eq. (3) is equal to $M$, which is shown in long-dashed black line. (Ref. \cite{47})}\label{Fig26}
\end{figure} 

This expression for the contrast is confirmed in measurements shown in Fig. 26. This expression is still valid when the size of measured transmission matrix $N^\prime$ is smaller than $N$ and the corresponding $M^\prime$ is correspondingly smaller than $M$. This is demonstrated by constructing a matrix of size $N^\prime=30$ from the measured transmission matrix of size $N = 66$ and calculating the contrast by phase conjugating the transmission matrix of size $N^\prime$. The contrast computed falls on the curve for $N = 30$ for different values of $M^\prime$. These results may be applied to measurements of the optical transmission matrix in which the size of the measured matrix $N^\prime$ is generally much smaller than $N$. In the limit $N\gg M$, the contrast approaches $M$. 

These results indicate that localized waves cannot be focused via phase conjugation because the value of $M$ is close to one. This is shown in Fig. 27 in which phase conjugation has been applied to focus the transmitted wave at the center of output surface for both diffusive and localized waves. Only for diffusive waves does a focal spot emerge from the background. We have recently demonstrated the use of phase conjugation to focus pulsed transmission through random media. By phase conjugating a time-dependent transmission matrix at a selected time delay, a pulse can be focused in space and time \cite{143}. 
\begin{figure}[htc]
\centering
\includegraphics[width=2in]{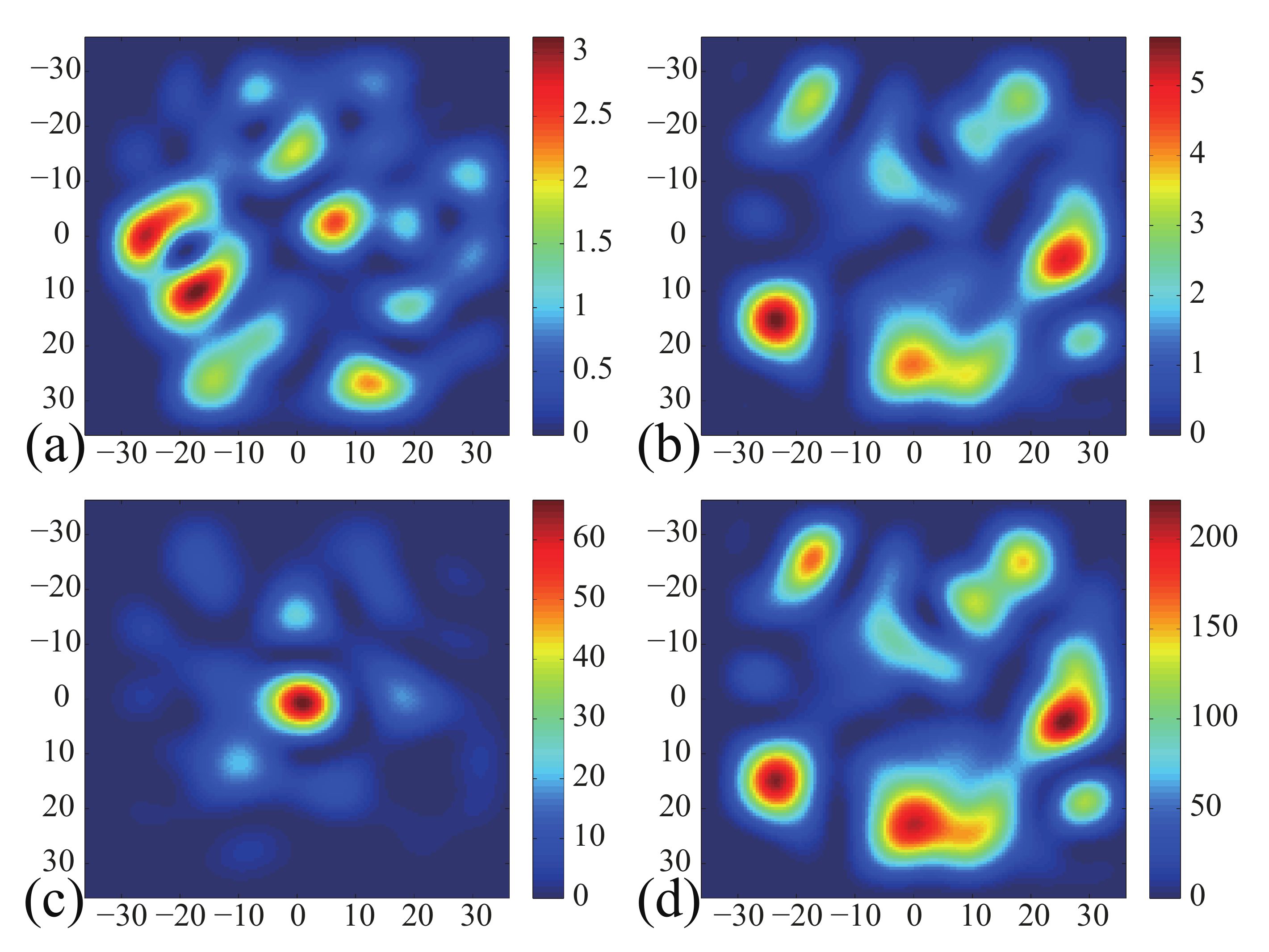}
\caption{Intensity speckle pattern generated for $L =23$ cm for diffusive waves (a) and for $L = 61$ cm for localized waves (b) normalized to the average intensity in the respective patterns. Focusing at the central point at the same frequency as in (a) and (b) via
phase conjugation is displayed in (c) and (d) with 66 and 30 input points, respectively. (Ref. \cite{48})}\label{Fig27}
\end{figure}

\section{Conclusion}
In this chapter, we have explored the mode and channel approaches to waves in random media. We believe that each of these approaches has the potential to provide a full description of transmission and its relation to the wave within the sample and that this will be of use in a wide variety of applications.

In recent work, we have considered four statistical characteristics of modes that have proven to be particularly promising for explaining steady-state and pulsed transmission and will be reported elsewhere. These characteristics are the statistics of the spacing and widths of modes, the degree of correlation in the speckle patterns of modes, and the mode transmittance. Correlation between speckle patterns includes correlation between the intensity patterns of modes as well as the average phase difference and the standard deviation of phase shift between these patterns. The mode transmittance represents the transmittance integrated over frequency for a particular mode and is obtained from a modal decomposition of the transmission matrix based on measurements of field transmission spectra between sets of points on the incident and output surfaces. The analysis of waves into modes is of particular interest in emission and lasing since it gives the density of states which is a key factor in the emission cross section as well as the lifetimes of modes. The relationship between modes and transmission eigenchannels can be elucidated by expressing the transmission eigenchannels at each frequency as the sum of projections of the eigenchannels of the transmission matrix for the individual modes upon the transmission eigenchannel \cite{145}. Precisely exciting a particular mode with a desired spatial distribution provides promise for control over energy deposition and collection within random media. 

The relationship between modes and transmission eigenchannels can be seen from the equality of the density of states obtained from the sum of the contributions of modes and of eigenchannels. The density of quasi-normal modes or resonances of a region per unit angular frequency, is the sum over Lorentzian lines, $\rho(\omega)=\frac{1}{\pi}\sum_n \frac{\Gamma_n/2}{(\Gamma_n/2)^2+(\omega-\omega_n)^2}$. This is found from the central frequencies and linewidths determined from a modal decomposition of fields at any points in the medium. The density of states can also be obtained from the sum of the contributions of each transmission eigenchannel, which are the derivatives with angular frequency of the composite phase shift of the eigenchannel, $\rho(\omega)=\frac{1}{\pi}\sum_n \frac{d\theta_n}{d\omega}$. The phase derivative is the intensity weighted phase derivative between all channels on the incident and output surfaces \cite{144}. $\frac{d\theta_n}{d\omega}$ is the transmission delay time for the $n^{th}$ transmission eigenchannel. When a complete measurement of the transmission matrix is made, $\frac{d\theta_n}{d\omega}$ is the integral of intensity inside the sample for the corresponding eigenchannel. The eigenchannel delay time and the associated intensity integral inside the sample increases with the transmission eigenvalue $\tau_n$. The density of states may be accurately measured from the transmission matrix as long as $N^\prime\gg M$. 

We have also explored the distribution of transmission eigenvalues and seen that in a particular transmission matrix, the statistics of relative transmission depend only upon $M$. The absolute distribution within a single matrix then depends upon these two parameters $M$ and $T$. Thus the distribution over a random ensemble of all transmission quantities depends only upon the joint distribution of $M$ and $T$. This represents a considerable simplification from the joint distribution of the full set of transmission eigenvalues $\tau_n$. Manipulation of the incident beam with knowledge of the transmission matrix makes it possible to achieve maximal focusing in a single transmission matrix with the peak intensity depending only upon $T$ and the contrast depending upon the value of $M$ in the measured matrix and the dimension of this matrix. Knowledge of the spectrum of both modes and channels may advance control over the wave projected within and through opaque samples for applications in imaging, and energy collection and delivery. 
\newline
\newline
{\bf Acknowledgments}

We would like to thank Jing Wang, Matthieu Davy, Patrick Sebbah, Valery Milner, Victor Kopp, Andrey Chabanov, Zhao-Qing Zhang, Xiaojun Cheng and Jerry Klosner for many stimulating discussions and for contributions to many of the results reviewed here. We thank the National Science Foundation for support under Grant Number DMR-1207446.

\end{document}